\documentclass[%
 aip,
 amsmath,amssymb,
 reprint,%
]{revtex4-1}

\newif\iflong
\longtrue

\iflong

\newcommand{\refappendix}[1]{\Cref{#1} of the SI}
\else

\newcommand{\refappendix}[2]{Section~{#2} of the SI}
\fi

\usepackage{graphicx}%
\usepackage{dcolumn}%
\usepackage{bm}%

\usepackage[utf8]{inputenc}
\usepackage[T1]{fontenc}
\usepackage{mathptmx}
\usepackage{etoolbox}

\usepackage[version=3]{mhchem}
\usepackage{xcolor}
\usepackage{amsmath}

\usepackage{amsthm}
\usepackage{physics}
\usepackage[caption=false]{subfig}
\usepackage[ruled,vlined,linesnumbered]{algorithm2e}

\usepackage{tikz}
\usepackage{xcolor}
\usepackage[export]{adjustbox}
\usepackage{paralist}
\usepackage{hyperref}
\usepackage{cleveref}

\usepackage[color=orange!60,textsize=scriptsize]{todonotes}

\DeclareSymbolFont{bbold}{U}{bbold}{m}{n}
\DeclareSymbolFontAlphabet{\mathbbold}{bbold}

\usepackage{float}
\usepackage{comment}
\usepackage{booktabs}

\iflong

\fi

\newif\ifmaintextcompiled
\maintextcompiledtrue

\makeatletter
\def\@email#1#2{%
 \endgroup
 \patchcmd{\titleblock@produce}
  {\frontmatter@RRAPformat}
  {\frontmatter@RRAPformat{\produce@RRAP{*#1\href{mailto:#2}{#2}}}\frontmatter@RRAPformat}
  {}{}
}%

\newcommand{\alghidebottomrule}{\renewcommand{\@algocf@post@ruled}}

\newcommand{\alghidetoprule}{\renewcommand{\@algocf@pre@ruled}}
\makeatother

\begin{document}

\title{Bosonic Condensed Phase Real-time Dynamics from Ring Polymer Molecular Dynamics}

\author{Yotam M. Y. Feldman}
\altaffiliation{These authors contributed equally to this work.}

\author{Sourav Karmakar}
\altaffiliation{These authors contributed equally to this work.}
\affiliation{School of Chemistry, Tel Aviv University, Tel Aviv 6997801, Israel.}
\affiliation{The Center for Computational Molecular and Materials Science, Tel Aviv University, Tel Aviv 6997801, Israel.}

\author{Sutirtha Paul}
\affiliation{Department of Physics and Astronomy, University of Tennessee, Knoxville, Tennessee 37996, USA.}

\author{Jacob Higer}
\affiliation{School of Physics, Tel Aviv University, Tel Aviv 6997801, Israel.}

\author{Adrian Del Maestro}
\affiliation{Department of Physics and Astronomy, University of Tennessee, Knoxville, Tennessee 37996, USA.}
\affiliation{Min H. Kao Department of Electrical Engineering and Computer Science, University of Tennessee, Knoxville, TN 37996, USA.}

\author{Barak Hirshberg}
\email[Corresponding author: ]{hirshb@tauex.tau.ac.il}
\affiliation{School of Chemistry, Tel Aviv University, Tel Aviv 6997801, Israel.}
\affiliation{The Center for Computational Molecular and Materials Science, Tel Aviv University, Tel Aviv 6997801, Israel.}

\begin{abstract}
We present a method for approximating real-time correlation functions and quantum transport coefficients of bosonic condensed phases. The direct evaluation of quantum real-time correlation functions in the path integral formulation is impossible due to a severe dynamical sign problem. Evaluating imaginary-time correlation functions and inverting them using analytic continuation is famously ill-posed. Ring polymer molecular dynamics (RPMD) provides an alternative approach resulting in accurate approximations for real-time correlation functions of condensed phases but neglects exchange effects. In this work, we develop a bosonic RPMD method, which is exact in the harmonic, short time, and high temperature limits. A critical enabling observation is that the Kubo-transformed correlation function for bosons is real only when the correlated observables are symmetric under exchange. We benchmark the method on harmonic and anharmonic model systems and then use it to directly obtain the Lieb-Liniger gas density-density real-time correlation functions at various temperatures and momentum transfers. This work enables the direct simulation of finite-temperature, real-time dynamics of large bosonic condensed phases using RPMD for the first time.
\end{abstract}

\maketitle

\newcommand{\TODO}[1]{\textcolor{red}{TODO: {#1}}}
\newcommand{\yotam}[1]{\textcolor{blue}{\bf {#1}}}
\newcommand{\jacob}[1]{\textcolor{magenta}{\bf {#1}}}
\newcommand{\notice}[1]{\textcolor{red}{\bf {#1}}}
\newcommand{\barak}[1]{\textcolor{purple}{\bf {#1}}}
\newcommand{\yotamsmall}[1]{{\footnotesize\color{magenta}[{\bf Yotam}: #1]}}
\newcommand{\alternative}[1]{\textcolor{magenta}{#1}}
\newcommand{\sourav}[1]{\textcolor{olive}{\bf {#1}}}

\newcommand{\set}[1]{\{{#1}\}}
\newcommand{\card}[1]{\left|{#1}\right|}
\newcommand{\sumcond}[2]{\substack{{#1}, \\ {#2}}}
\newcommand{\bigO}{\mathcal{O}}

\newcommand{\fact}[1]{{#1}!}
\newcommand{\pfact}[1]{\fact{\left(#1\right)}}
\newcommand{\Symset}[1]{\mathcal{S}({#1})}
\newcommand{\Symfromto}[2]{\mathcal{S}[{#1},{#2}]}
\newcommand{\SymN}[1]{\mathcal{S}[1,{#1}]}

\newcommand{\permsubgroup}[3]{\mathcal{S}_#1 \left(#2 \rightsquigarrow #3\right)}

\newcommand{\cycleof}[2]{c_{#1}({#2})}
\newcommand{\nextof}[3]{{#3}({#1})={#2}}
\newcommand{\prevparticle}[1]{{#1}^{-}}
\newcommand{\nextparticle}[1]{{#1}^{+}}
\newcommand{\particleup}{v}
\newcommand{\particledown}{u}
\newcommand{\cyclenotate}[1]{({#1})}

\newcommand{\pos}{{\bf R}}
\newcommand{\posbead}[2]{{\bf r}_{#1}^{#2}}
\newcommand{\beadpos}[2]{\posbead{#1}{#2}}
\newcommand{\mass}{m}

\newcommand{\springfrequency}{\omega_P}
\newcommand{\springconstant}{\mass \springfrequency^2}
\newcommand{\springenergyprefix}{\frac{1}{2} \springconstant}

\newcommand{\rdiffsquared}[4]{\left(\beadpos{#1}{#2} - \beadpos{#3}{#4}\right)^2}
\newcommand{\interparticleforce}[1]{\springenergyprefix \sum_{j=1}^{P-1}{\rdiffsquared{#1}{j+1}{#1}{j}}}

\newcommand{\repsym}{G}
\newcommand{\rep}[1]{\repsym[{#1}]}

\newcommand{\boltzmann}[1]{e^{-\beta {#1}}}

\newcommand{\Vfromto}[2]{V_{\text{PBC}}^{[{#1},{#2}]}}
\newcommand{\Vfrom}[1]{\Vfromto{#1}{N}}
\newcommand{\Vto}[1]{\Vfromto{1}{#1}}
\newcommand{\Vall}{\Vfromto{1}{N}}

\newcommand{\originalpotential}[1]{\Vto{#1}}
\newcommand{\originalpotentialnotation}[1]{V_B^{({#1})}}

\newcommand{\Eperm}[1]{E^{#1}}
\newcommand{\Efromto}[2]{E_{\text{PBC}}^{[{#1},{#2}]}}
\newcommand{\Enk}[2]{\Efromto{{#1}-{#2}+1}{{#1}}}
\newcommand{\Einterior}[1]{E_{\textit{int}}^{({#1})}}
\newcommand{\permutationthree}[3]{\left(\begin{smallmatrix} 
1 & 2 & 3\\
{#1} & {#2} & {#3}
\end{smallmatrix}\right)}

\newcommand{\snip}[2]{\rep{#1} \textit{ snips at } {#2}}
\newcommand{\fullsnip}[3]{\rep{#1} \textit{ snip from } {#2} \textit{ to } {#3}}
\newcommand{\project}[2]{{#1}_{| {#2}}}

\newcommand{\beadderive}[3]{\beadderiveexplicit{#1}{#2}{#3}}
\newcommand{\beadderiveexplicit}[3]{\nabla_{\beadpos{#1}{#2}} {#3}}
\newcommand{\beadforce}[3]{- \beadderive{#1}{#2}{#3}}

\newcommand{\Prrep}{\Prerepperm{\sigma}}
\newcommand{\Prerepperm}[1]{\Pr(\rep{#1})}
\newcommand{\Prrepnext}[2]{\Pr\left(\nextof{#1}{#2}{\rep{\sigma}}\right)}
\newcommand{\Prrepnextpair}[4]{\Pr\left(\nextof{#1}{#2}{\rep{\sigma}}, \nextof{#3}{#4}{\rep{\sigma}}\right)}

\newcommand{\windingcap}{\mathcal{W}}
\newcommand{\windingcapvector}{\windingcap}
\newcommand{\winding}[2]{\mathbf{w}_{#1}^{#2}}
\newcommand{\rdiffwinding}[4]{\beadpos{#1}{#2} + \winding{#1}{#2} L - \beadpos{#3}{#4}}
\newcommand{\rdiffsquaredwinding}[4]{\left(\rdiffwinding{#1}{#2}{#3}{#4}\right)^2}
\newcommand{\windingsumindex}[2]{\substack{{\winding{#2}{1}=-{#1},\ldots,{#1}} \\ {\ldots} \\ {\winding{#2}{P}=-{#1},\ldots,{#1}}}}
\newcommand{\windingsequence}{\set{\mathbf{w}}}
\newcommand{\windingsequencefromto}[2]{\set{\mathbf{w}}^{[{#1},{#2}]}}
\newcommand{\rdiffsquaredwindingcrazy}[4]{\left(\beadpos{#1}{#2} \overset{\windingcap}{-} \beadpos{#3}{#4}\right)^2}
\newcommand{\localspringenergysym}{\mu}
\newcommand{\edgewinding}[4]{\localspringenergysym(\beadpos{#1}{#2},\beadpos{#3}{#4})}
\newcommand{\edgewindingspecificdistinguishable}[2]{\edgewindingspecific{#1}{#2}{#1}{{#2}+1}}
\newcommand{\edgewindingspecific}[4]{\localspringenergysym(\beadpos{#1}{#2},\beadpos{#3}{#4},\winding{#1}{#2})}
\newcommand{\edgewindingdistinguishable}[2]{\sum_{\winding{#1}{#2}}{\edgewindingspecificdistinguishable{#1}{#2}}}
\newcommand{\Prereppermwinding}[2]{\Pr(\rep{#1}, {#2})}
\newcommand{\Epermdistinguishable}[1]{\Eperm{#1}}
\newcommand{\Prwinding}[2]{\Pr\left(\winding{#1}{#2}\right)}
\newcommand{\Prwindingpair}[4]{\Pr\left(\winding{#1}{#2}, \winding{#3}{#4}\right)}
\newcommand{\Prrepandwinding}[3]{\Pr\left(\nextof{#1}{#2}{\rep{\sigma}}, \winding{#1}{#3}\right)}
\newcommand{\Prcond}[2]{\Pr\left(#1 \mid #2\right)}
\newcommand{\Prwindingcond}[3]{\Prcond{\winding{#1}{#3}}{\nextof{#1}{#2}{\rep{\sigma}}}}
\newcommand{\Prwindingcondwinding}[4]{\Prcond{\winding{#1}{#2}}{\winding{#3}{#4}}}
\newcommand{\windingmic}[2]{(\winding{#1}{#2})^{\ast}}
\newcommand{\Prwindingmic}[2]{\Pr\left(\windingmic{#1}{#2}\right)}

\newcommand{\efromtodbeta}[2]{A^{[{#1}, {#2}]}}
\newcommand{\Einteriordbeta}[1]{A_{\textit{int}}^{({#1})}}

\newcommand{\Prwindingext}[4]{
\frac{
\edgewindingspecific{#1}{#2}{#3}{#4}
}
    {\edgewinding{#1}{#2}{#3}{#4}}
}
\newcommand{\Prwindingextshort}[4]{
\frac{\edgewindingspecific{#1}{#2}{#3}{#4}}
    {
    \left(
    \sum_{\winding{#1}{#2}=-\windingcap}^{\windingcap}{\edgewindingspecific{#1}{#2}{#3}{#4}}
    \right)
    }
}
\newcommand{\Prwindingextshortdistinguishable}[2]{
\frac{\edgewindingspecificdistinguishable{#1}{#2}}
    {\edgewinding{#1}{#2}{#1}{{#2}+1}}
}
\newcommand{\Epermorig}[1]{\Eperm{#1}}
\newcommand{\Vtoorig}[1]{V^{[1,{#1}]}}
\newcommand{\Efromtoorig}[2]{E^{[{#1},{#2}]}}
\newcommand{\Enkorig}[2]{\Efromtoorig{{#1}-{#2}+1}{{#1}}}

\newcommand{\beadposdim}[3]{({r}_{#1}^{#2})_i}
\newcommand{\posbeaddim}[3]{\beadposdim{#1}{#2}{#3}}
\newcommand{\windingcomponent}[3]{\left({w}_{#1}^{#2}\right)_{#3}}
\newcommand{\windingdim}[3]{\windingcomponent{#1}{#2}{i}}
\newcommand{\rdiffsquaredwindingdim}[5]{\left(\beadposdim{#1}{#2}{#5} + \windingdim{#1}{#2}{#5} L - \beadposdim{#3}{#4}{#5}\right)^2}

\newcommand{\Epermidorig}{\Epermorig{\textit{id}}}
\newcommand{\Vdist}{V_{\text{D}}}
\newcommand{\Predistpermwinding}[2]{\Pr({#2})}

\newcommand{\unit}[2]{#1\;\text{#2}}
\newcommand{\density}[1]{#1\;\text{\AA}^{-3}}
\newcommand{\length}[1]{#1\;\text{\AA}}
\newcommand{\massunit}[1]{#1\;\text{u}}
\newcommand{\temperature}[1]{#1\;\mathrm{K}}
\newcommand{\dt}[1]{#1\;\text{fs}}
\newcommand{\meV}[1]{\unit{#1}{meV}}

\newcommand{\addprime}[1]{{#1}^{\prime}}
\newcommand{\iprime}{\addprime{i}}
\newcommand{\jprime}{\addprime{j}}
\newcommand{\kprime}{\addprime{k}}
\newcommand{\uprime}{\addprime{u}}
\newcommand{\vprime}{\addprime{v}}
\newcommand{\ellprime}{\addprime{\ell}}
\newcommand{\sigmaprime}{\addprime{\sigma}}

\newcommand{\permop}[1]{\hat{P}_{#1}}

\newcommand{\allbeadpositions}{\mathbf{R}}
\newcommand{\allbeadmomenta}{\mathbf{p}}
\newcommand{\kubocorrdist}[2]{\tilde{C}^{(D)}_{{#1}{#2}}}
\newcommand{\kubocorrbosons}[2]{\tilde{C}^{(B)}_{{#1}{#2}}}
\newcommand{\hamiltonian}{\hat{H}}
\newcommand{\classicalhamiltonian}{H}
\newcommand{\opA}{\hat{A}}
\newcommand{\opB}{\hat{B}}
\newcommand{\position}{\mathbf{r}}
\newcommand{\positiononedim}{x}
\newcommand{\opPosition}{\hat{\position}}
\newcommand{\rpmdcorrdist}[2]{\tilde{K}^{(D)}_{{#1}{#2}}}
\newcommand{\rpmdopdist}[1]{{#1}^{(D)}_P}
\newcommand{\beadpostimedist}[3]{{\beadpos{#1}{#2}}^{(D)}({#3})}
\newcommand{\rphamiltonian}{\classicalhamiltonian^{({D})}_P}
\newcommand{\bosonshamiltonian}{\classicalhamiltonian^{({B})}_P}
\newcommand{\rpmdcorrbosons}[2]{\tilde{K}^{(B)}_{{#1}{#2}}}
\newcommand{\rpmdopbosons}[1]{{#1}^{(B)}_P}
\newcommand{\beadpostimebosons}[3]{{\beadpos{#1}{#2}}^{(B)}({#3})}
\newcommand{\symmetricpositionop}{\hat{Q}}

\newcommand{\partitionfuncdistinguishable}{Z^{(D)}}
\newcommand{\partitionfuncbosons}{Z^{(B)}}
\newcommand{\partitionfuncpathintegralbosons}{Z_P^{(B)}}
\newcommand{\beadposbosons}[2]{\posbead{#1}{#2 \, (B)}}
\newcommand{\momentumbead}[2]{{\bf p}_{#1}^{#2}}
\newcommand{\beadmomentumdist}[2]{\momentumbead{#1}{#2}}
\newcommand{\beadmomentumbosons}[2]{\momentumbead{#1}{#2 \, (B)}}

\newcommand{\compos}{\textbf{r}_c}
\newcommand{\commomentum}{\textbf{p}_c}

\newcommand{\symmetricbasis}{\mathcal{S}}
\newcommand{\generalbasis}{\mathcal{U}}
\newcommand{\symbasisperp}{\symmetricbasis^{\perp}}

\newcommand{\tracebosons}{\Tr_S}

\section{Introduction}
Simulating the finite-temperature quantum dynamics of condensed phases remains a key outstanding challenge of computational chemistry and physics.~\cite{zhang_recent_2016} In particular, real-time correlation functions are of interest since they determine important response properties through fluctuation--dissipation relations.~\cite{kubo_statistical-mechanical_1957, kubo_fluctuation-dissipation_1966} 
For example, the velocity and dipole autocorrelation functions determine the diffusion coefficient~\cite{miller_quantum_2005, miller_quantum_2005-1, markland_quantum_2008} and IR spectrum,~\cite{rossi_how_2014} respectively. 
Density-density correlations determine the dynamic structure factor (DSF), an important quantity measurable via neutron scattering, in the study of superfluids and other bosonic and fermionic phases.~\cite{PhysRevB.97.184520,DelMaestro2022,myung_prediction_2022} 

Methods based on the path integral formulation of quantum mechanics, such as Path Integral Molecular Dynamics (PIMD)~\cite{parrinello_study_1984} and Monte Carlo (PIMC),~\cite{ceperley_path_1995} are some of the most powerful techniques for simulating bosonic condensed phases at thermal equilibrium. They are based on the exact mapping of the quantum system to an extended phase space of classical ring polymers (see details below). However, their extension to real-time properties suffers from the infamous dynamical sign problem.~\cite{makri_feynman_1991} The state-of-the-art solution of simulating PIMD or PIMC to evaluate imaginary-time correlation functions is possible, but analytic continuation is required to obtain real-time correlation functions and transport properties.~\cite{gallicchio_calculation_1996, krilov_real_1999, rabani_quantum_2000, rabani_calculation_2002, golosov_analytic_2003, habershon_quantum_2007} Despite important recent advancements,~\cite{yoon_analytic_2018, dornheim_ab_2018, fei_nevanlinna_2021, fournier_artificial_2020, DelMaestroAnalyticContinuation2022, ADMaestroSciPostPhysCodeb.39} analytic continuation is fundamentally highly numerically unstable, with small amounts of noise in the input imaginary-time data leading to large errors in the predicted properties. 

For real-time correlation functions of distinguishable particles (obeying Boltzmann statistics), an alternative to analytic continuation is available. Approximate path integral approaches such as centroid molecular dynamics (CMD)~\cite{cao_new_1993, cao_formulation_1994, cao_formulation_1994-1, cao_formulation_1994-2, cao_formulation_1994-3} and ring polymer molecular dynamics (RPMD)~\cite{craig_quantum_2004, habershon_ring-polymer_2013} use the classical real-time dynamics in the extended phase space of ring polymers to approximate the quantum Kubo-transformed real-time correlation function. They have emerged in recent years as powerful tools for obtaining quantum response properties of solids and liquids.~\cite{miller_quantum_2005, miller_quantum_2005-1, craig2006inelastic, habershon_ring-polymer_2013, castro2025vibrational, markland_nuclear_2018, musil2022quantum, TreninsRossiPRL2025, althorpe_path-integral_2021} However, these techniques are based on path integral molecular dynamics which ordinarily neglects exchange effects.
Although bosonic CMD simulations have been previously~%
attempted, they were limited to small systems,~\cite{roy_feynman_1999-1, blinov_operator_2001, kinugawa_path_1999, kinugawa_path_2001, kinugawa_semiclassical_2001} due to the exponential scaling incurred by summing over all the permutations of the identical particles.
Including all permutations in large bosonic systems is also a challenge for semiclassical initial-value representation methods. 
To the best of our knowledge, the method of Nakayama
and Makri,~\cite{Nakayama2005} who combined PIMC permutation sampling with forward-backward semiclassical dynamics to compute the dynamic structure factor of superfluid helium, has been the only available approach for describing real-time bosonic dynamics. However, this method may suffer from numerical instability~%
due the dynamical sign problem.~\cite{Nakayama2004,Chen2010} It was also used only in the high-momentum-transfer regime, approximating the bosonic correlation function by a sum over single-particle contributions.~\cite{Nakayama2004,Nakayama2005}

We recently developed an efficient PIMD algorithm that includes bosonic exchange effects exactly, reducing the computational scaling with the number of particles from exponential to quadratic.~\cite{hirshberg2019path,quadratic-pidmb} We then expanded this approach to fermions~\cite{doi:10.1063/5.0008720} and extended systems,~\cite{higer2025PBC} higher-order propagators~\cite{Higer_Suzuki}, and used it to predict a supersolid phase of deuterium at high pressure.~\cite{myung_prediction_2022} However, bosonic PIMD simulations have so far been limited to evaluating static quantum properties at thermal equilibrium. 

In this paper, we develop a method to obtain dynamics and transport properties of bosonic systems using RPMD for the first time. We derive a bosonic RPMD estimator for the Kubo-transformed real-time correlation function between two operators that incorporates exchange symmetry and is exact in the high temperature, short time and harmonic limits, like standard RPMD. 
Our method is therefore the first PIMD-based quantum dynamics approach that can be efficiently applied to bosonic condensed phase systems.
We derive the bosonic RPMD estimator and prove the limits in which it is exact. 
We benchmark our method on trapped bosons and find that it provides an accuracy roughly on par with the performance of RPMD for distinguishable particles.

We then apply the method to the Lieb-Liniger gas~\cite{Lieb-Liniger1963}, an experimentally realizable \cite{Paredes2004TonksGirardeau,Kinoshita2004Observation,Kinoshita2005, Kinoshita2006QuantumNewtonsCradle,vanAmerongen2008YangYang,Haller2009SuperTonks} system of one-dimensional bosons with contact interactions, and obtain its real-time density-density correlations for various temperatures and momentum transfers. We find that our method provides real-time correlation functions and dynamic structure factors that are in a very good agreement with exact results obtained from the ABACUS algorithm.~\cite{CauxCalabresePRA2006, PanfilCauxPRA2014, MeinertPRL2015, Granet-EsslerSciPost2020}
Specifically, we find that bosonic RPMD agrees with the exact results better than analytic continuation at small to moderate momentum transfers. 
Our method can therefore be used to obtain real-time, condensed-phase, finite-temperature dynamics directly and with no ad-hoc parameters or numerical instabilities. 
At high momentum transfers, RPMD performs less well than analytic continuation due to the non-linearity of the density operator.

The paper is organized as follows. \Cref{sec:RPMD-theory} reviews the RPMD theory for distinguishable particles. The extension of RPMD to bosonic systems is introduced in \Cref{sec:RPMDB-theory}, and \Cref{sec:importance-of-symmetry} describes the importance of operator symmetry in the Kubo-transformed bosonic correlation function. In \Cref{sec:results}, we present the results. We first derive the exact limits of bosonic RPMD in \Cref{sub:exact_limits_for_bosonic_ring_polymer_correlation_functions}. We then apply bosonic RPMD to model systems in \Cref{sub:numerical_results-models} and to the Lieb-Liniger gas in \Cref{sub:LL-results}. We conclude in \Cref{sec:conclusions}.

\section{Theory}

\subsection{Background: Kubo-transformed and RPMD correlation functions}
\label{sec:RPMD-theory}
We consider a Hamiltonian of the form $\hamiltonian = \frac{1}{2\mass} \sum_{\ell=1}^N {\bf \hat{p}}^2_\ell + U({\bf \hat{r}}_1,...,{\bf \hat{r}}_N)$.
The quantum Kubo-transformed correlation function~\cite{kubo2012statistical} is the central object in path integral approximations to real-time quantum dynamics.~\cite{Althorpe2021} For distinguishable particles, denoted by the superscript $(D)$, it can be written as
\begin{equation}
\label{eq:kubo-transformed-correlation}
    \kubocorrdist{A}{B}(t) = \frac{1}{\beta \partitionfuncdistinguishable}\int_{0}^{\beta}{d\lambda \, \Tr{e^{-(\beta-\lambda)\hamiltonian} \opA(0) e^{-\lambda \hamiltonian} \opB(t)}},
\end{equation}
where $\partitionfuncdistinguishable = \Tr{e^{-\beta \hamiltonian}}$ is the canonical partition function, and $\opB(t) = e^{+it\hamiltonian/\hbar} \opB e^{-it\hamiltonian/\hbar}$ is the Heisenberg-evolved operator. 
The Kubo-transformed correlation function's adequacy for classical approximations is underscored by its symmetry properties: for $\opA,\opB$ that are position-dependent Hermitian operators, $\kubocorrdist{A}{B}(t)$ is a real, even function of $t$. Additionally, it is symmetric in the order of operators, $\kubocorrdist{A}{B}(t) = \kubocorrdist{B}{A}(t)$.~\cite{craig_quantum_2004}

In RPMD~\cite{craig_quantum_2004} of distinguishable particles, the Kubo-transformed correlation function of~\Cref{eq:kubo-transformed-correlation} is approximated by a time correlation function in a classical, ``isomorphic'' system. This system is defined by the ring polymer Hamiltonian for distinguishable particles,
\begin{equation}
\label{eq:distinguishable-hamiltonian}
    \begin{split}
    \rphamiltonian = 
        &\sum_{\ell=1}^{N}{
        \sum_{j=1}^{P}{
            \frac{({\beadmomentumdist{\ell}{j})}^2}{2m}
        }}
        +
        \springenergyprefix
        \sum_{\ell=1}^{N}{
        \sum_{j=1}^{P}{
             \rdiffsquared{\ell}{j}{\ell}{j+1}
        }}
        \\
        &+
        \sum_{j=1}^{P}{
            \frac{1}{P} U(\beadpos{1}{j},\ldots,\beadpos{N}{j})
        }
    \end{split}
\end{equation}
In~\Cref{eq:distinguishable-hamiltonian}, $\beadpos{\ell}{j}$ is the position of the $j$th bead of particle $\ell$, and $\beadmomentumdist{\ell}{j}$ is the conjugate momentum. The beads of each particle are connected through harmonic springs of frequency $\springfrequency = \frac{\sqrt{P}}{\beta \hbar}$ to rings, such that $\beadpos{\ell}{P+1} = \beadpos{\ell}{1}$. For each $j = 1, \dots, P$, beads sharing the same index $j$ interact through the scaled physical potential.

RPMD then approximates the quantum Kubo correlation function through the RPMD correlation function. For two position-dependent operators $\opA,\opB$ it is defined by
\begin{equation}
\label{eq:ring-polymer-correlation}
    \rpmdcorrdist{A}{B}(t) = \ev{ \rpmdopdist{A}(0) \rpmdopdist{B}(t)}_{\rphamiltonian},
\end{equation}
where $\ev{\dots}_{\rphamiltonian}$ denotes the canonical ensemble phase space average corresponding to the classical ring-polymer Hamiltonian $\rphamiltonian$. 
In~\Cref{eq:ring-polymer-correlation}, each operator is averaged over the imaginary-time slices (bead indices),
\begin{equation}
\label{eq:ring-polymer-estimator}
    \rpmdopdist{A}(t) = \frac{1}{P} \sum_{j=1}^{P}{A(\beadpostimedist{1}{j}{t},\ldots,\beadpostimedist{N}{j}{t}}),
\end{equation}
and its time-dependence comes from the propagation of bead positions $\beadpostimedist{\ell}{j}{t}$ according to Hamilton's equations (constant energy) for $\rphamiltonian$.%

Like the Kubo-transformed correlation function, the RPMD correlation function $\rpmdcorrdist{A}{B}(t)$ is a real, even function of time, and $\rpmdcorrdist{A}{B}(t) = \rpmdcorrdist{B}{A}(t)$.~\cite{craig_quantum_2004} As an approximation to $\kubocorrdist{A}{B}$, it is known to be exact at three important limits:~\cite{craig_quantum_2004} 
(1) the short time limit $\rpmdcorrdist{A}{B}(0) = \kubocorrdist{A}{B}(0)$;
(2) the high temperature limit, $\rpmdcorrdist{A}{B}(t) = \kubocorrdist{A}{B}(t)$ as $\beta \to 0$;
(3) the harmonic limit, i.e, when $U$ is a harmonic trap, in which case the autocorrelation between two linear, position-dependent operators is exact at all times $t$, e.g., $\rpmdcorrdist{\position}{\position}(t) = \kubocorrdist{\position}{\position}(t)$.

So far, RPMD always assumed Boltzmann statistics. 
The quantum linear response theory of bosonic systems uses the bosonic Kubo-transformed correlation function $\kubocorrbosons{A}{B}$, which is written exactly as in~\Cref{eq:kubo-transformed-correlation}, except all traces are performed over a properly symmetrized basis:
\begin{equation}
\label{eq:kubo-transformed-correlation-bosonic}
    \kubocorrbosons{A}{B}(t) = \frac{1}{\beta \partitionfuncbosons}\int_{0}^{\beta}{d\lambda \, \tracebosons\Big\{e^{-(\beta-\lambda)\hamiltonian} \opA(0) e^{-\lambda \hamiltonian} \opB(t)}
    \Big\},
\end{equation}
where $\tracebosons$ indicates that the trace is taken exclusively within the sub-space of completely symmetric wavefunctions.
We next present a method to approximate such real-time correlation functions for bosons using RPMD.

\subsection{Bosonic RPMD correlation functions}\label{sec:RPMDB-theory}

In this section, we introduce bosonic RPMD correlation functions as a means to approximate bosonic Kubo-transformed real-time correlation functions.
Considering the two elements in the definition of distinguishable RPMD correlation functions (\Cref{eq:ring-polymer-correlation,eq:ring-polymer-estimator}), we obtain the bosonic
variant by modifying the Hamiltonian governing the dynamics of the classical ring polymers, and demonstrate that the same observable estimators remain applicable as long as the operators used in the correlation are symmetric under particle exchange.

To begin, we replace the classical ring polymer Hamiltonian for distinguishable particles $\rphamiltonian$ by the classical ring polymer Hamiltonian for bosonic particles that we previously used to obtain static properties through PIMD:~\cite{hirshberg2019path, quadratic-pidmb} 
\begin{equation}
\label{eq:bosons-ring-polymer-hamiltonian}
    \bosonshamiltonian = 
        \sum_{\ell=1}^{N}{
        \sum_{j=1}^{P}{
            \frac{({\beadmomentumdist{\ell}{j})}^2}{2m}
        }}
        + \Vtoorig{N} + \sum_{j=1}^{P}{\frac{1}{P} U(\beadpos{1}{j},\ldots,\beadpos{N}{j})}
        ,
\end{equation}
where the bosonic spring potential is defined recursively by
\begin{align}
\boltzmann{\Vtoorig{N}} &= \frac{1}{N} \sum_{k=1}^{N}{\boltzmann{\left(\Vtoorig{N-k} + \Enkorig{N}{k}\right)}}
\\
\boltzmann{\Vtoorig{0}} &= 1
\\
\Enkorig{N}{k} &= \springenergyprefix \sum_{\ell=N-k+1}^{N}{\sum_{j=1}^{P}{\rdiffsquared{\ell}{j+1}{\ell}{j}}},
\label{eq:Enk}
\end{align}
and in~\Cref{eq:Enk}, $\beadpos{\ell}{P+1} = \beadpos{\ell+1}{1}$ except for $\beadpos{N}{P+1} = \beadpos{N-k+1}{1}$. 
See Ref.~\citenum{quadratic-pidmb} for the full details.
The merit of this Hamiltonian is that it provides exact sampling in the static limit $t=0$, since its classical partition function coincides with the quantum bosonic partition function at $P \to \infty$. Furthermore, it can be evaluated efficiently,~\cite{quadratic-pidmb} in time that is quadratic in $N$ and linear in $P$.
Importantly, new to this work, for $t>0$, we also propagate the estimators using $\bosonshamiltonian$, ensuring that the dynamics preserve the bosonic quantum equilibrium distribution.

Next, we determine the appropriate estimators for the two observables in the classical RPMD correlation function. The same estimators employed for distinguishable particles (\Cref{eq:ring-polymer-estimator}) can be used for bosons, under the condition that the operators $\opA,\opB$ are symmetric under exchange, i.e., $A(\opPosition_{\sigma(1)}, \ldots, \opPosition_{\sigma(N)}) = A(\opPosition_1, \ldots, \opPosition_N)$ for every permutation $\sigma$ of the $N$ identical particles, and likewise for $\opB$. As we will show soon, the symmetry of $\opA,\opB$ is not a technical limitation, but a necessary requirement for semiclassical approximations to the bosonic Kubo-transformed quantum correlation function.

We define the bosonic RPMD correlation function of two symmetric position-dependent operators $\opA,\opB$ by 
\begin{equation}\label{eq:rpmdb-corrfn}
    \rpmdcorrbosons{A}{B}(t) = \ev{
    \rpmdopbosons{A}(0) \rpmdopbosons{B}(t)}_{\bosonshamiltonian},
\end{equation}
where $\ev{\dots}_{\bosonshamiltonian}$ denotes the canonical ensemble phase space average corresponding to the classical ring-polymer Hamiltonian $\bosonshamiltonian$, and $\rpmdopbosons{A}(t)$ is defined by
\begin{equation}
\label{eq:ring-polymer-estimator-bosons}
    \rpmdopbosons{A}(t) = \frac{1}{P} \sum_{j=1}^{P}{A(\beadpostimebosons{1}{j}{t},\ldots,\beadpostimebosons{N}{j}{t}}),
\end{equation}
and similarly for $\rpmdopbosons{B}(t)$. The bead positions $\beadpostimebosons{\ell}{j}{t}$ evolve in time according to Hamilton's equations for the bosonic ring polymer Hamiltonian $\bosonshamiltonian$ (\Cref{eq:bosons-ring-polymer-hamiltonian}).

We show in~\Cref{sub:exact_limits_for_bosonic_ring_polymer_correlation_functions} that, with these definitions, bosonic RPMD correlation functions satisfy the same symmetry properties and exactness conditions as the ordinary RPMD correlation functions for distinguishable particles. 
As a precursor, we discuss the importance of the symmetry of $\opA,\opB$ for the symmetry properties of the bosonic Kubo-transformed correlation function itself.

\subsection{The importance of operator symmetry}
\label{sec:importance-of-symmetry}

\begin{figure}[ht]
\centering
\includegraphics[width=0.5\textwidth]{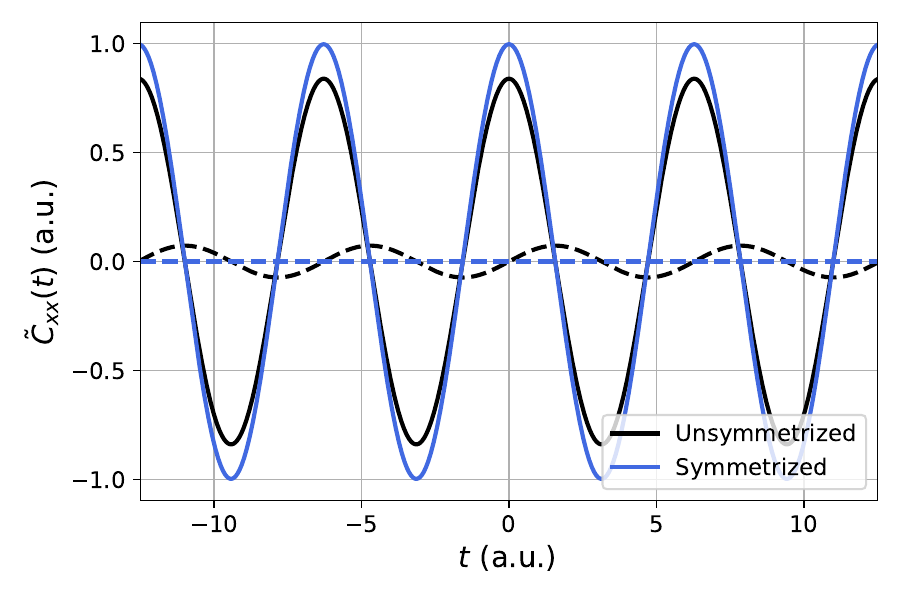}
\caption{Exact Kubo-transformed correlation functions of $N=2$ non-interacting bosons in a one-dimensional harmonic trap at $\beta = \unit{1}{a.u.}$. Solid and dashed lines represent the real and imaginary part of the correlation function, respectively. For the unsymmetrized position operator (in black), $\tilde{C}_{xx}(t)\equiv\tilde{C}_{x_1x_1}(t)$. For the symmetrized position operator (in blue), $\tilde{C}_{xx}(t) \equiv \frac{1}{2}\tilde{C}_{x_1+x_2,x_1+x_2}(t)$.}
\label{fig:kubo-symmetry-importance-symmetric}
\end{figure}

The symmetry of $\opA,\opB$ is important not only for our bosonic RPMD correlation functions, but is also fundamental for the exact quantum correlations. If $\opA,\opB$ are not symmetric under exchange, the Kubo-transformed correlation function $\kubocorrbosons{A}{B}$ is not necessarily real or symmetric. For example, \Cref{fig:kubo-symmetry-importance-symmetric} 
shows the exact single-particle position autocorrelation $\tilde{C}_{x_1x_1}(t)$
in a system of $N=2$ bosons, with a mass of 1 atomic unit (a.u.), in a one-dimensional harmonic trap ($\hbar \omega = \unit{1}{a.u.}$) at $\beta = \unit{1}{a.u.}$. 
Seven harmonic oscillator energy states were enough to obtain a converged result.
The resulting unsymmetrized correlation function is complex, and the imaginary part is an odd function of $t$. In contrast, for the symmetrized operator, it is a real and even function of $t$.
We show that when $\opA,\opB$ are symmetric, Hermitian, position-based operators, the bosonic Kubo-transformed correlation function is a real, even function of time, and symmetric w.r.t.\ the order of $\opA,\opB$ (\refappendix{sec:bosonic_kubo_transformed_correlation_functions_symmetry_properties}). These properties make the Kubo-transformed correlation function amenable to semiclassical approximation. 

The symmetry of $\opA,\opB$ is required for them to be an observable property of indistinguishable particles; our insight is that this is necessary also for the correlation function to be well-behaved. Put differently, it is impossible to distinguish between self-correlations  
and cross-correlations 
between indistinguishable particles. 
In practice, many properties of interest in bosonic condensed phases, such as the conductivity and the dynamic structure factor, correspond to correlations of symmetric observables.
Single-particle correlations,
like the single-particle velocity autocorrelation (used to define the tracer diffusion coefficient~\cite{10.1093/9780191947971.001.0001}), and the incoherent part of the density-density correlation function (used to define the incoherent dynamic structure factor), do not satisfy this criterion.

\section{Results} \label{sec:results}
\subsection{Exact limits for bosonic RPMD correlation functions}
\label{sub:exact_limits_for_bosonic_ring_polymer_correlation_functions}
We now discuss the exact limits of bosonic RPMD correlation functions. It is clear from the definition that $\rpmdcorrbosons{A}{B}$ is a real function since $\opA$ and $\opB$ are both real. Other properties merit extended discussion.

\paragraph{Short time limit.}
As $t \to 0$, the bosonic RPMD correlation function becomes exact, $\rpmdcorrbosons{A}{B}(t) \to \kubocorrbosons{A}{B}(t)$. This relies on the bosonic ring-polymer Hamiltonian (\Cref{eq:bosons-ring-polymer-hamiltonian}) producing the exact bosonic statistics in the static ($t=0$) case,~\cite{hirshberg2019path,quadratic-pidmb} and the exchange symmetry of $\opA,\opB$ which justifies the average over the imaginary-time slices in $\rpmdopbosons{A},\rpmdopbosons{B}$ (\Cref{eq:ring-polymer-estimator-bosons}). 
A full derivation appears in~\refappendix{sec:bosonic_short_time_limit}.

\paragraph{Parity and symmetry.}
$\rpmdcorrbosons{A}{B}$ is an even function of time and symmetric with respect to exchanging $\opA,\opB$. The argument proceeds analogously to that for distinguishable particles (see~\refappendix{sec:even_symmetric_function}).

\paragraph{High temperature limit.}

In the $\beta \to 0$ limit, the harmonic springs become very stiff, making the ring polymers shrink into single beads. Consequently, the bosonic RPMD estimator, as its distinguishable counterpart, reduce to a classical correlation function. 

\paragraph{Harmonic limit.}
For a system of non-interacting bosons in a harmonic trap, the RPMD correlation function exactly captures the center of mass  autocorrelation,

\begin{equation}
\label{eq:harmonic-limit-bosons}
    \frac{1}{N} \rpmdcorrbosons{\sum_{\ell=1}^{N}{\positiononedim_\ell}}{, \sum_{\ell=1}^{N}{\positiononedim_\ell}}(t) = \frac{1}{N}\kubocorrbosons{\sum_{\ell=1}^{N}{\positiononedim_\ell}}{, \sum_{\ell=1}^{N}{\positiononedim_\ell}}(t),
\end{equation}
where the position operators for the $N$ bosons are $\hat{\positiononedim}_1,\ldots,\hat{\positiononedim}_N$.
This is the analog of the result for distinguishable particles, where (in the absence of interactions) the RPMD correlation function is exact for single-particle autocorrelations,
\begin{equation}
\rpmdcorrdist{\positiononedim_\ell}{\positiononedim_\ell}(t) = \kubocorrdist{\positiononedim_\ell}{\positiononedim_\ell}(t).
\end{equation}
The proof of the harmonic limit in the distinguishable case relies on showing that the centroid of each ring evolves classically; in our case, bosonic exchange leads to $N$-body correlations, and hence only the centroid of all the beads of all particles combined is shown to evolve classically.
The proof of~\Cref{eq:harmonic-limit-bosons} appears in~\refappendix{sec:bosonic_harmonic_limit}. 

\subsection{Numerical results} \label{sub:numerical_results-models}
We first apply bosonic RPMD to a system of non-interacting bosons in one-dimensional harmonic, weakly anharmonic, and strongly anharmonic traps. These model potentials were previously used to benchmark RPMD for distinguishable particles.\cite{craig_quantum_2004} We compare the bosonic RPMD real-time autocorrelation functions with the exact results for those model systems.
We then compute the dynamic structure factor of the Lieb-Liniger model\cite{Lieb-Liniger1963, Lieb1963} where exact results are available. 
All the Kubo-transformed real-time correlation functions shown hereafter are normalized by the number of particles. Computational details are given in~\refappendix{sec:simulation-details}.

\begin{figure}[ht]
\centering
\includegraphics[width=0.5\textwidth]{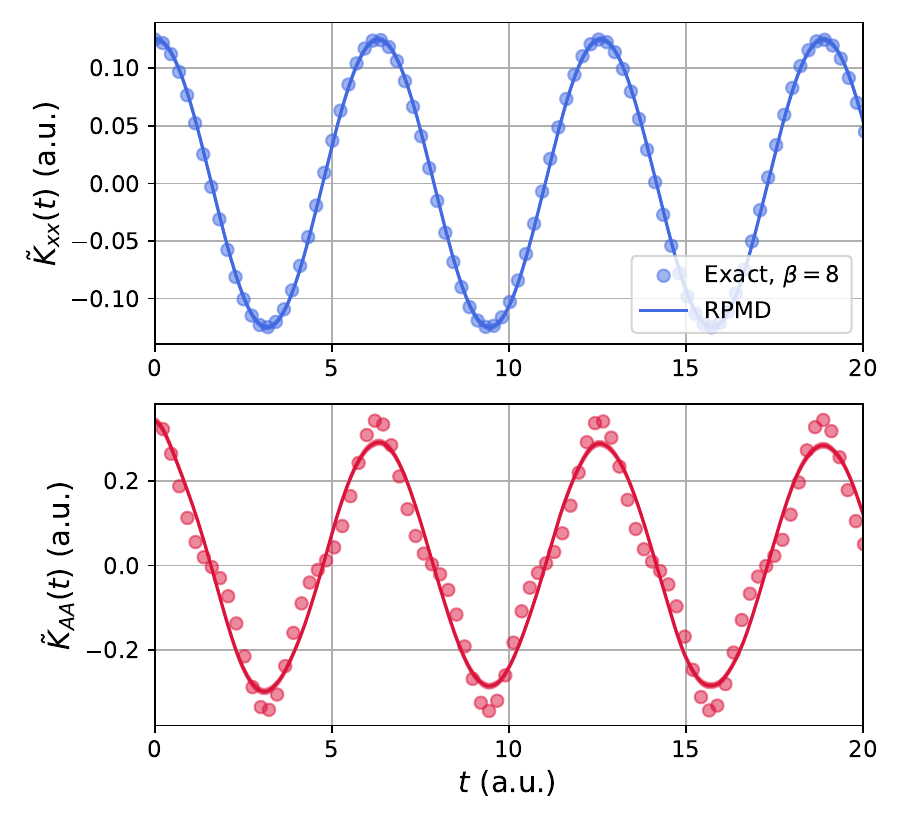}
\caption{Bosonic RPMD autocorrelation functions for the position operator $\hat{x} = \sum_{\ell=1}^N \hat{x}_\ell$ (upper panel) and the nonlinear operator $\hat{A} = \sum_{\ell=1}^N \hat{x}^3_\ell$ (lower panel) for a system of $N=4$ non-interacting bosons in a harmonic trap at $\beta =\unit{8}{a.u.}$ Note that the position coordinate is expressed in units of the harmonic oscillator length $\sqrt{\hbar / m \omega}$. Symbols represent the exact Kubo-transformed autocorrelation function.
}
\label{fig:rpmdb-sho}
\end{figure}

\subsubsection{Model potentials}
\paragraph{Harmonic potential}
We first apply bosonic RPMD to a system of $N=4$ non-interacting bosons with mass $m = \unit{1}{a.u.}$ in a harmonic trap, $V(x)=\frac{1}{2} m \omega^2 x^2$, with $\hbar \omega = \unit{3}{meV}$, and compute the bosonic RPMD correlation function for two position-dependent observables: $\hat{x} = \sum_{\ell=1}^N \hat{x}_\ell$ and $\hat{A} = \sum_{\ell=1}^N \hat{x}^3_\ell$. \Cref{fig:rpmdb-sho} shows the bosonic RPMD correlation function at $\beta = \unit{8}{a.u.}$. Since bosonic RPMD is exact for linear position-dependent observables in the harmonic limit (\Cref{sub:exact_limits_for_bosonic_ring_polymer_correlation_functions}), the position autocorrelation function is identical to the analytical result. However, for the nonlinear operator $\hat{A}$, bosonic RPMD is not exact at all times, even when the trap is harmonic. Bosonic RPMD reproduces the short time behavior correctly, as expected, as well as approximates the dynamics reasonably well at longer times. This finding confirms that the accuracy of bosonic RPMD decreases with increasing nonlinearity of the observable,
as is the case in RPMD of distinguishable particles.~\cite{craig_quantum_2004}

\begin{figure}[ht]
\centering
\includegraphics[width=0.5\textwidth]{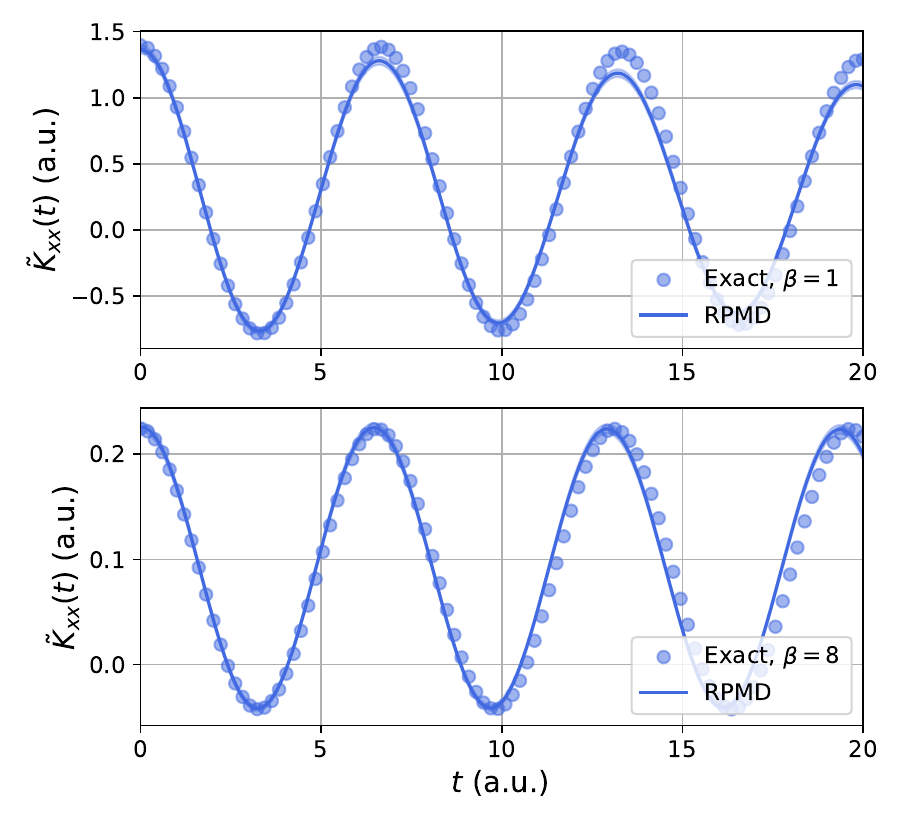}
\caption{Bosonic RPMD position autocorrelation function for a system of $N=4$ non-interacting bosons in a weakly anharmonic trap at $\beta= \unit{1}{a.u.}$ (upper panel) and $\beta=\unit{8}{a.u.}$ (lower panel). Symbols represent the exact Kubo-transformed correlation function obtained numerically.}
\label{fig:rpmdb-weakanh}
\end{figure}

\paragraph{Weakly anharmonic potential}
We now explore the validity of bosonic RPMD beyond the harmonic limit. To this end, we consider a weakly anharmonic potential and a strongly anharmonic potential (a quartic trap), previously used to benchmark RPMD for distinguishable particles.~\cite{craig_quantum_2004} The weakly anharmonic trap is of the form $V(x) = \frac{x^2}{2} + \frac{x^3}{10} + \frac{x^4}{100}$.~\Cref{fig:rpmdb-weakanh} shows a comparison between the bosonic RPMD and the numerically exact Kubo-transformed autocorrelation function (computed by diagonalizing the Hamiltonian) of the position operator $\hat{x}$ for a system of $N=4$ non-interacting bosons with  mass $m = \unit{1}{a.u.}$ at two temperatures, $\beta=\unit{1}{a.u.}$ and $\unit{8}{a.u.}$ Bosonic RPMD shows excellent agreement with the exact results at both temperatures. At high temperature ($\beta = \unit{1}{a.u.}$), as the system explores more of the anharmonic domain, the bosonic RPMD autocorrelation function deviates from the exact values at long times.
Conversely, at low temperature ($\beta = \unit{8}{a.u.}$), the system mostly explores the potential minimum, so bosonic RPMD remains accurate even at longer timescales. Comparable accuracy is also found in the case of RPMD for distinguishable particles.~\cite{craig_quantum_2004}

\begin{figure}[ht]
\centering
\includegraphics[width=0.475\textwidth]{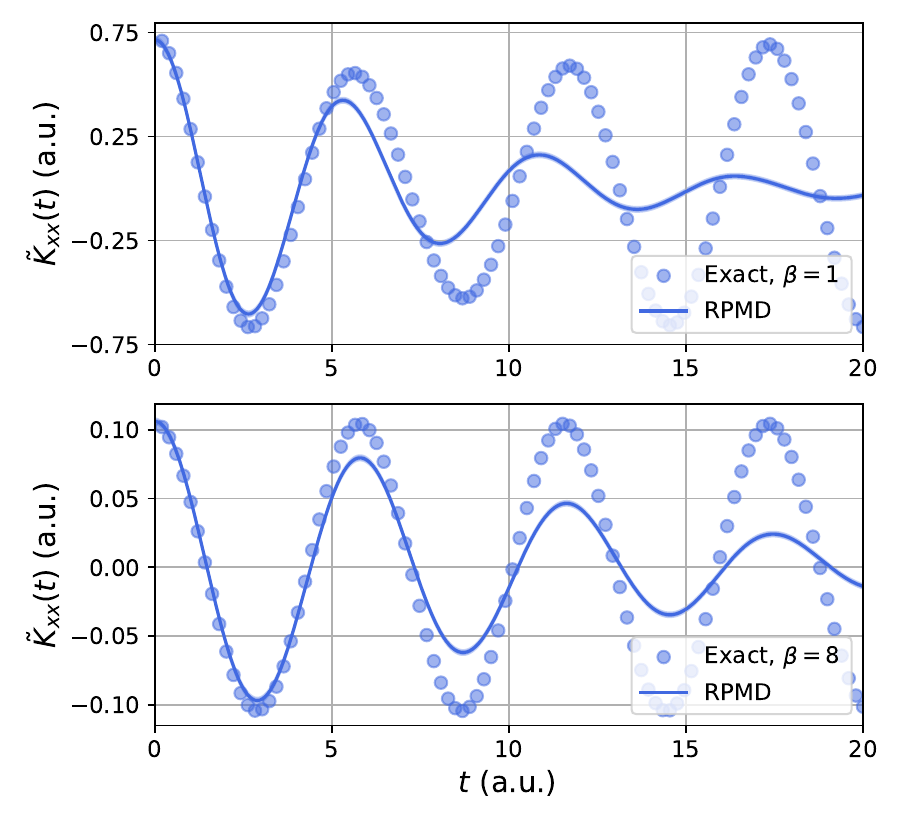}
\caption{Bosonic RPMD position autocorrelation function for a system of $N=4$ non-interacting bosons in a quartic trap at $\beta=\unit{1}{a.u.}$ (upper panel) and $\beta=\unit{8}{a.u.}$ (lower panel). Symbols represent the exact Kubo-transformed autocorrelation function obtained numerically.}
\label{fig:rpmdb-quartic}
\end{figure}

\paragraph{Strongly anharmonic potential}
The strongly anharmonic potential is a quartic trap of the form $V(x)=\frac{1}{4}x^4$.~\Cref{fig:rpmdb-quartic} shows the bosonic RPMD position autocorrelation function and its comparison with the numerically exact Kubo-transformed correlation function for a system of $N=4$ non-interacting bosons with mass $m = \unit{1}{a.u.}$ at two temperatures, $\beta=\unit{1}{a.u.}$ and $\unit{8}{a.u.}$. Bosonic RPMD accurately captures the short time behavior of the position autocorrelation function at both temperatures.
In particular, at low temperature, bosonic RPMD captures quantum oscillations even at long times. However, as the timescale increases and quantum dynamical effects become more pronounced, bosonic RPMD begins to deviate from the exact quantum correlation function. Importantly, the accuracy of bosonic RPMD is comparable to that of distinguishable-particle RPMD.~\cite{craig_quantum_2004}

\subsection{Application to the Lieb-Liniger Gas}\label{sub:LL-results}
We now apply our bosonic RPMD method to a system of $N$ interacting bosons in a one-dimensional box of length $L$ with periodic boundary conditions 
described by the Lieb-Liniger Hamiltonian\cite{Lieb-Liniger1963, Lieb1963} of the form
\begin{equation}
\label{eq:LL-ham}
\hat{H}_{\text{LL}} = 
-\frac{\hbar^2}{2m}
\sum_{\ell=1}^{N}{
    \pdv[2]{x_{\ell}}
}
+ g_{\text{LL}}
\sum_{\ell>s}^{N}{
    \delta\left(
        x_\ell - x_s
    \right)
},
\end{equation}
where $m$ is the mass and $g_{\text{LL}}$ is the interaction strength. 
We consider a repulsive interaction $g_{\text{LL}} >0$, which corresponds to the scattering length $a_{\text{LL}} = -\frac{2\hbar^2}{mg_{\text{LL}}}$.
This model is of interest because it is experimentally realizable~\cite{Paredes2004TonksGirardeau,Kinoshita2004Observation,Kinoshita2005, Kinoshita2006QuantumNewtonsCradle,vanAmerongen2008YangYang,Haller2009SuperTonks} and exhibits interesting physics due to the combination of strong repulsion and confinement to one dimension.
It is also convenient because exact solutions exist for the Lieb-Liniger gas.
At $T=0$, the energy spectrum can be obtained exactly using the Bethe ansatz method.~\cite{bethe1931theorie} For finite temperature, the thermal Bethe ansatz~\cite{Yang-Yang1969, Yang1970} has been developed to obtain the thermodynamic quantities of this system. 
Several works have also been devoted to obtaining the dynamical quantities.~\cite{CauxCalabresePRA2006, PanfilCauxPRA2014, MeinertPRL2015, Granet-EsslerSciPost2020, Granet2021, Li_2023, Senese2026, Panfil2026}

Working with a Dirac delta interaction is impractical for RPMD, so we instead approximate the pair interaction using a Gaussian, which is a finite and smooth function of the inter-particle distance. We therefore employ the Hamiltonian
\begin{equation}\label{eq:modified-LL-ham}
\hat{H} = 
-\frac{\hbar^2}{2m}
\sum_{\ell=1}^{N}{ 
    \pdv[2]{x_{\ell}}
}
+
\frac{
    g
}{
    \sqrt{2\pi \sigma_g^2}
}
\sum_{\ell>s}^{N}{
    e^{
        -\frac{
            (x_\ell - x_s)^2
        }{
            2\sigma_g^2
        }
    }
}.
\end{equation}
For a sufficiently small $\sigma_g$, we expect to recover the essential physics of the Lieb-Liniger model (\Cref{eq:LL-ham}), which depend mostly on the scattering length.
Because the two-body scattering length of a Gaussian interaction can be estimated,~\cite{jeszenszki2018s} for a given Lieb-Liniger coupling $g_{\text{LL}}$, we determine the parameters $\set{g, \sigma_g}$ by matching the scattering length of the two models. Details of the parameters can be found in~\refappendix{sec:parameters-mLL}. We simulate the model using a bosonic RPMD algorithm with periodic boundary conditions.~\cite{higer2025PBC, higerpimdbGitHub} 

We aim to compute the density-density correlation function, also known as the intermediate scattering function (ISF), for the Lieb-Linger model directly in real time. 
The bosonic ISF is defined as
\begin{equation}\label{eq:isf}
F(k, t)
= \frac{1}{NZ^{(B)}}
\tracebosons\left\{
    \boltzmann{
        \hat{H}
    }
    \hat{\rho}_k \hat{\rho}_{-k}\left(t\right)
\right\},
\end{equation}
where $\hbar k$ denotes the momentum imparted to the system, for instance in a neutron scattering process.
The density operator in momentum space, $\hat{\rho}_k = \sum_{\ell=1}^N e^{-i k\hat{x}_\ell}$, is the Fourier transform of the real space density operator, $\hat{\rho} (x) = \sum_{\ell=1}^N \delta(x_\ell - x)$.

The bosonic Kubo-transformed ISF is then
\begin{equation}\label{eq:kubo-isf}
\tilde{F}(k, t)=\frac{1}{\beta N Z^{(B)}}\int_0^{\beta} d\lambda 
\tracebosons\left\{
    e^{-(\beta - \lambda) \hat{H}} \hat{\rho}_k e^{-\lambda \hat{H}}\hat{\rho}_{-k}(t)
\right\}.
\end{equation}
Using \Cref{eq:rpmdb-corrfn}, we can write the RPMD approximation to the Kubo-transformed ISF as
\begin{equation}
\label{eq:density-corr-rpmd}
\tilde{K}_{\rho_k \rho_{-k}}(t) = 
\ev{
    \rho_{k, P}(0) \rho_{-k, P}(t)
}_{\bosonshamiltonian},
\end{equation}
where
\begin{equation}
\rho_{k, P}(t) = \frac{1}{P} \sum_{j=1}^P \sum_{\ell=1}^N e^{-ikx_\ell^j (t)},
\end{equation}
is the RPMD estimator for the density operator $\hat{\rho}_k$.

\begin{figure}[ht]
\centering
\includegraphics[width=.5\textwidth]{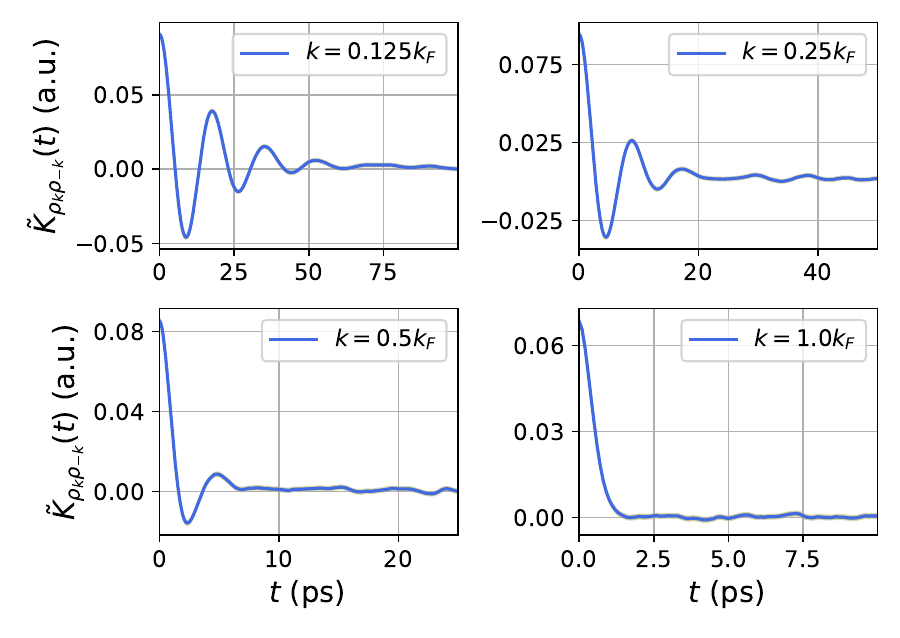}
\caption{Bosonic RPMD density-density real-time correlation function for a system of $N=32$ interacting bosons at $T=\temperature{1}$ for four different momentum transfers $k / k_F = 0.125, 0.25, 0.5, 1.0$.}

\label{fig:isf-LL}
\end{figure}

\Cref{fig:isf-LL} shows the real-time RPMD density-density correlation function for a system of $N = 32$ interacting bosons of mass $m = \unit{48.48}{amu}$ in a 1D box of length $L = \length{16}$ at $T = \temperature{1}$. 
The Gaussian interaction parameters considered here are $g = \unit{6}{K\AA}$ and $\sigma_g = \length{0.118234}$, corresponding to $g_{\text{LL}}=\unit{10}{K\AA}$. For small momentum transfer, $\tilde{K}_{\rho_k \rho_{-k}}(t)$ exhibits oscillations that persist at long times. With increasing momentum transfer, $\tilde{K}_{\rho_k \rho_{-k}}(t)$ decays faster. 

The Fourier transform of the Kubo ISF, $\tilde{S}(k,\omega)=\frac{1}{2\pi}\int\limits_{-\infty}^{\infty} dt \tilde{F}(k,t) e^{-i\omega t}$, is related to the DSF as
\begin{equation}
\label{eq:dsf-from-isf}
S \left(
    k, \omega
\right)
=
\frac{
    \beta \hbar \omega
}{
    1-e^{-\beta \hbar \omega}
} 
\tilde{S} \left(
    k, \omega
\right).
\end{equation}
To obtain the DSF, we perform the Fourier transform of $\tilde{K}_{\rho_k \rho_{-k}}(t)$ and multiply the spectrum by the Kubo-factor $\beta \hbar \omega (1-e^{-\beta \hbar \omega})^{-1}$, following~\Cref{eq:dsf-from-isf}. 

\begin{figure}[ht]
\centering
\includegraphics[width=.5\textwidth]{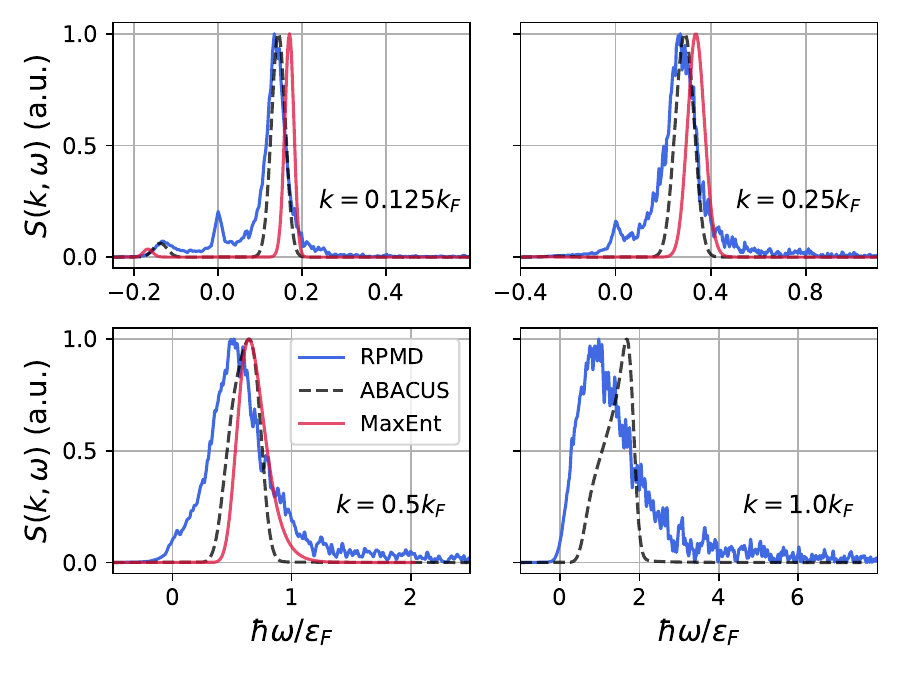}
\caption{Dynamic structure factor (scaled) for four different momentum transfers considered in~\Cref{fig:isf-LL} where $\varepsilon_F = \hbar^2 k_F^2/2m$. Prior to computing the dynamic structure factor, we multiply the Kubo-transformed real-time density-density correlation function with a decaying exponential to suppress the fluctuations at long times (see~\refappendix{sec:noise_suppresion_dsf} for details). The dashed lines denote results obtained from the ABACUS algorithm for the Lieb-Liniger Hamiltonian. MaxEnt data correspond to the analytic continuation results for the system in \Cref{eq:modified-LL-ham}.}

\label{fig:dsf-LL}
\end{figure}

The DSF obtained from $\tilde{K}_{\rho_k \rho_{-k}}(t)$ is shown in~\Cref{fig:dsf-LL}, in comparison to the ABACUS algorithm,~\cite{caux2009ABACUS} which is numerically exact. Note that the ABACUS predictions use the Hamiltonian of \Cref{eq:LL-ham}, not \Cref{eq:modified-LL-ham}, but otherwise use the same simulation parameters. We also employ a smoothening procedure using the two particle level spacing. Overall, for both methods, we find that the peak position in the DSF shifts towards higher frequency with increasing momentum transfer, since $\tilde{K}_{\rho_k \rho_{-k}}(t)$ decays faster. 

\begin{figure}[ht]
\centering
\includegraphics[width=.5\textwidth]{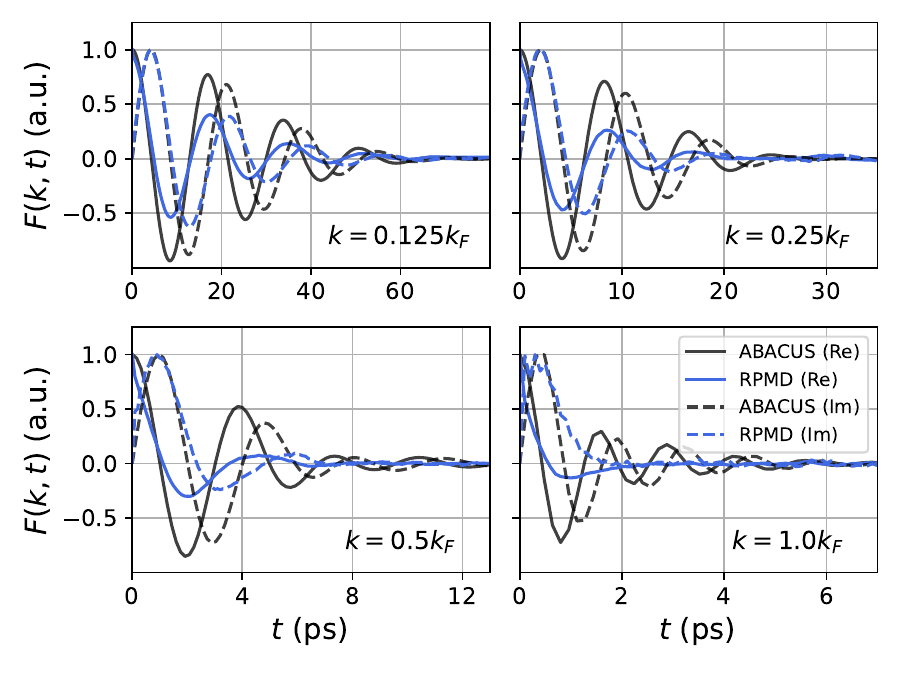}
\caption{Intermediate scattering function, or the real-time density-density correlation function (scaled), for four different momentum transfers considered in~\Cref{fig:isf-LL,fig:dsf-LL}, obtained by performing an inverse Fourier transform of the DSF in~\Cref{fig:dsf-LL}.}

\label{fig:acf-LL}
\end{figure}

Since bosonic RPMD is a better approximation for linear operators, it is expected to reproduce the DSF better for small momentum transfers, i.e., $\frac{k}{k_F}=0.125, 0.25$ ($k_F = \pi\frac{N}{L}$ is the Fermi wavevector arising due to refermionization of bosons with hard cores in 1D).~\cite{Girardeau1960} We indeed find excellent agreement with the ABACUS result for this regime. RPMD recovers the correct peak location and width, as well as the appearance of a small peak at negative frequencies for small $k$.
The bosonic RPMD DSF is only slightly broader compared to ABACUS.  
To understand these differences, we plot the ISF obtained by performing the inverse Fourier transform of the two DSF in~\Cref{fig:acf-LL}. We find that the bosonic RPMD real-time correlation function decays faster, which leads to the slight width over-estimation.
Finally, we note the presence of a low-intensity peak at $\omega=0$ in the DSF for bosonic RPMD, which does not appear in ABACUS. As shown in~\refappendix{sec:comparison-boson-dist-LL}, the peak at $\omega=0$ does not exist for distinguishable particles. We hypothesize this peak originates from the diffusion of the ring polymers involving partial particle-exchanges (see~\refappendix{sec:comparison-boson-dist-LL} for details).

For higher momentum transfers, the density operator becomes strongly nonlinear, and bosonic RPMD becomes less accurate. As a result, the DSF for bosonic RPMD shows a larger deviation both in the peak location and width, as shown in~\Cref{fig:dsf-LL} for $\frac{k}{k_F}=0.5, 1.0$. The broader DSF is again due to the faster overall decay in RPMD in the corresponding correlation function~(\Cref{fig:acf-LL}). 
For both momentum transfers, we find comparably good agreement at several other temperatures (see~\refappendix{sec:isf_diff_Ts} for details).

 An alternative route of obtaining the DSF is performing analytic continuation of the imaginary-time ISF computed from PIMC~\cite{ceperley_path_1995,Boninsegni:2006ed,Boninsegni:2006gc,pimcrepo} for the same  Hamiltonian in \Cref{eq:modified-LL-ham}. While it is a famously ill-posed problem, we perform the analytic continuation using the maximum entropy algorithm~\cite{Jarrell1996,Bergeron2016} in the ACFlow library \cite{Huang2023,Huang2026}. For small momentum transfers, we find that the peak positions are shifted from the ABACUS and RPMD results. For high momentum  transfers, analytic continuation agrees better than RPMD with the exact results, which is expected due to the operator non-linearity.
 
\section{Conclusions}
\label{sec:conclusions}
In this work, we presented bosonic RPMD, a method to compute Kubo-transformed real-time correlation functions of bosonic condensed phases. Bosonic RPMD is exact in the short time, high temperature and harmonic limits for linear operators. It also incorporates bosonic statistics exactly, successfully capturing the effect of the exchange symmetry on real-time correlation function (see~\refappendix{comparison-boson-dist-modelpotential} for a detailed comparison).
We benchmarked bosonic RPMD on non-interacting bosons in harmonic and anharmonic traps. For the harmonic case, we confirmed the method is exact and for the weakly anharmonic potential, it also captured the long time behavior accurately. For the strongly anharmonic potential, bosonic RPMD correctly reproduced the short time behavior of the real-time correlation function. 

We then applied bosonic RPMD to the Lieb-Liniger gas. We directly computed the finite-temperature density-density correlation functions in real time at various temperatures (see~\refappendix{sec:isf_diff_Ts} for details) and momentum transfers. We also obtained from them the dynamic structure factor. 
For momentum transfers $k/k_F=0.125,0.25$, RPMD gives very good approximations of the exact real-time correlation functions. It reproduces the peak location and width accurately. For momentum transfers $k/k_F=0.5,1.0$, bosonic RPMD captures the peak location slightly worse, and overestimates the width, likely due to the non-linear nature of the density operator at high $k$.
We did not apply bosonic RPMD to higher momentum transfers, since this regime can be handled by complementary methods that decompose the correlation function into single-particle contributions~\cite{Nakayama2004, Nakayama2005}.
Comparing bosonic RPMD with analytic continuation of the imaginary time density-density correlation function computed via PIMC, we find that RPMD better reproduces the location peak for small momentum transfers and that analytic continuation performs better for high momentum transfers. 

Bosonic RPMD enables the computation of the real-time quantum correlation functions in large bosonic condensed phases for the first time. The method opens the door to studying real-time dynamics of trapped ultracold gases, superfluid helium under confinement, and other strongly correlated bosonic phases. It does not suffer from the dynamical sign problem and does not rely on ill-posed analytic continuation. It can also be used as a prior to improve analytic continuation, as previously demonstrated for distinguishable particles.~\cite{habershon_quantum_2007}

\bibliography{refs, new-refs}%

\iflong
\onecolumngrid
\clearpage
\appendix
\section{Proofs}

\subsection{Bosonic Kubo-transformed correlation functions: symmetry properties}
\label{sec:bosonic_kubo_transformed_correlation_functions_symmetry_properties}

We repeat the derivation of Ref.~\citenum{craig2006a} for distinguishable particles, while identifying the points in the proof where the exchange symmetry properties of $\opA$ and $\opB$ are required.

Let $\hat{\Pi}_{\sigma}$ be the permutation operator corresponding to the permutation $\sigma$. We begin by noting that the symmetrizer,
\begin{equation}
    \hat{S} 
    = 
    \frac{
        1
    }{
        \fact{N}
    }
    \sum_{\sigma}{
        \hat{\Pi}_{\sigma}
    },
\end{equation}
commutes with the Hamiltonian, which means they share a common eigenbasis of the full $N$-particle Hilbert space. This basis, which we denote as $\set{\ket{n}}$, consists of eigenstates belonging to the totally symmetric subspace,
and eigenstates belonging to its orthogonal complement.
The basis of the former will be denoted as $\symmetricbasis$ while the basis of the latter will be denoted as $\symbasisperp$. In particular, the resolution of identity in the full Hilbert space can be written as
\begin{equation}
    \hat{\mathbb{I}} 
    = 
    \sum_{n \in \symmetricbasis}{
        \ket{n}\bra{n}
    }
    +
    \sum_{m \in \symbasisperp}{
        \ket{m}\bra{m}
    }.
\end{equation}
Expanding the bosonic Kubo-transformed correlation function in $\symmetricbasis$,
\begin{align}
    \kubocorrbosons{A}{B}(t)
    &= \frac{1}{\beta \partitionfuncbosons}
        \int_{0}^{\beta}{d\lambda \, 
        \sum_{n \in \symmetricbasis}{
            e^{-\beta E_n} e^{\lambda E_n} 
            \mel{
                n
            }{
                \opA e^{-\lambda \hamiltonian} e^{+it\hamiltonian/\hbar} \opB
            }{
                n
            }
            e^{-it E_n/\hbar}
        }
        }.
    \\
    \intertext{Using the previous resolution of the identity:}
    &=
        \frac{1}{\beta \partitionfuncbosons}
        \int_{0}^{\beta}d\lambda \, 
        \sum_{n \in \symmetricbasis}{
            e^{-\beta E_n} e^{\lambda E_n} 
            \bra{
                n
            }
            \opA e^{-\lambda \hamiltonian} e^{+it\hamiltonian/\hbar}
            \left[
            \sum_{m \in \symmetricbasis}
            \ket{  
                m
            }
            \mel{
                m    
            }{
                \opB
            }{
                n
            }
            +
            \sum_{k \in \symbasisperp}
            \ket{  
                k
            }
            \mel{
                k    
            }{
                \opB
            }{
                n
            }
            \right]
            e^{-it E_n/\hbar}
        }
    \\
    \intertext{Recall that $\opB$ is a symmetric operator, so $\opB\ket{n}$ is a totally symmetric state, which implies that $\mel{k}{\opB}{n} = 0$. Therefore,}
    &= \frac{1}{\beta \partitionfuncbosons}
        \int_{0}^{\beta}{d\lambda \, 
        \sum_{n \in \symmetricbasis}{
        \sum_{m \in \symmetricbasis}{
            e^{-\beta E_n} e^{\lambda E_n} e^{-\lambda E_m} e^{+it E_m/\hbar} \mel{n}{\opA}{m} \mel{m}{\opB}{n} e^{-it E_n/\hbar}
        }
        }
        }.
    \\
    \intertext{Integrating over the $\lambda$-dependent terms yields the expression (whose exact form is inspired by Ref.~\citenum{10.1093/9780191947971.001.0001})
    }
    &= \frac{1}{\beta \partitionfuncbosons}
        \sum_{n \in \symmetricbasis}{
        \sum_{m \in \symmetricbasis}{
            e^{-it(E_n - E_m)/\hbar} \mel{n}{\opA}{m} 
                                     \mel{m}{\opB}{n}
                                     \frac{e^{-\beta E_m} - e^{-\beta E_n}}{E_n - E_m}
        }}.
\end{align}

The rest of the proof follows as in the case of distinguishable particles:
\begin{enumerate}
    \item $\kubocorrbosons{B}{A}(t) = \kubocorrbosons{A}{B}(-t)$ by interchanging the sums:
        \begin{align}
            \kubocorrbosons{B}{A}(t) &=
            \frac{1}{\beta \partitionfuncbosons}
                \sum_{n \in \symmetricbasis}{
                \sum_{m \in \symmetricbasis}{
                    e^{-it(E_n - E_m)/\hbar} \mel{n}{\opB}{m}
                                             \mel{m}{\opA}{n}
                                             \frac{e^{-\beta E_m} - e^{-\beta E_n}}{E_n - E_m}
                }
                }
            \\
            &= \frac{1}{\beta \partitionfuncbosons}
                \sum_{m \in \symmetricbasis}{
                \sum_{n \in \symmetricbasis}{
                    e^{-i(-t)(E_m - E_n)/\hbar} \mel{m}{\opA}{n}
                                             \mel{n}{\opB}{m}
                                             \frac{e^{-\beta E_n} - e^{-\beta E_m}}{E_m - E_n}
                }
                }
            \\
            &=
            \kubocorrbosons{A}{B}(-t).
        \end{align}
        
    \item $\kubocorrbosons{A}{B}(t)^{*} = \kubocorrbosons{A}{B}(t)$ by interchanging the sums and using the fact that $\opA,\opB$ are Hermitian:
        \begin{align}
            \kubocorrbosons{A}{B}(t)^{*} &=
            \frac{1}{\beta \partitionfuncbosons}
                \sum_{n \in \symmetricbasis}{
                \sum_{m \in \symmetricbasis}{
                    e^{+it(E_n - E_m)/\hbar} \mel{m}{\opA^{\dagger}}{n}
                                             \mel{n}{\opB^{\dagger}}{m}
                                             \frac{e^{-\beta E_m} - e^{-\beta E_n}}{E_n - E_m}
                }
                }
            \\
            &= \frac{1}{\beta \partitionfuncbosons}
                \sum_{m \in \symmetricbasis}{
                \sum_{n \in \symmetricbasis}{
                    e^{-it(E_m - E_n)/\hbar} \mel{m}{\opA}{n}
                                             \mel{n}{\opB}{m}
                                             \frac{e^{-\beta E_n} - e^{-\beta E_m}}{E_m - E_n}
                }
                }
            \\
            &=
            \kubocorrbosons{A}{B}(t).
        \end{align}

    \item $\kubocorrbosons{A}{B}(t)^{*} = \kubocorrbosons{A}{B}(-t)$ by the fact that we can choose the eigenstates $|n\rangle$ of a real Hamiltonian to be real functions in position-basis, and so the matrix elements of the position-dependent operators $\opA,\opB$ are real as well, and we have $\mel{n}{\opA}{m} = \mel{m}{\opA}{n}$ (and likewise for $\opB$):
        \begin{align}
            \kubocorrbosons{A}{B}(t)^{*} &=
            \frac{1}{\beta \partitionfuncbosons}
                \sum_{n \in \symmetricbasis}{
                \sum_{m \in \symmetricbasis}{
                    e^{+it(E_n - E_m)/\hbar} \mel{n}{\opA}{m} 
                                             \mel{m}{\opB}{n}
                                             \frac{e^{-\beta E_m} - e^{-\beta E_n}}{E_n - E_m}
                }
                }
            \\
            &=
            \kubocorrbosons{A}{B}(-t).
        \end{align}
\end{enumerate}

The symmetry properties $\kubocorrbosons{A}{B}(t) = \kubocorrbosons{A}{B}(-t)$, $\kubocorrbosons{A}{B}(t) = \kubocorrbosons{B}{A}(t)$ follow from the combination of the above properties~\cite{craig_quantum_2004}, which include also the property $\kubocorrbosons{A}{B}(t)^{*} = \kubocorrbosons{A}{B}(t)$.

\subsection{Bosonic short time limit}
\label{sec:bosonic_short_time_limit}
Starting from the exact expression for the bosonic Kubo-transformed correlation function with the trace expanded in position basis,
\begin{align}
\kubocorrbosons{A}{B}(0) &= \frac{1}{\beta \partitionfuncbosons}\int_{0}^{\beta}{d\lambda \, 
    \int{
    	\frac{1}{\fact{N}}\sum_{\sigma}{\langle \, \position_1,\ldots,\position_N \, | \, e^{-(\beta-\lambda)\hamiltonian} \opA e^{-\lambda \hamiltonian} \opB \, | \, \position_{\sigma(1)}, \ldots, \position_{\sigma(N)} \rangle \, d\position_1 \ldots d\position_N}
    	}
    },
\\
\intertext{we approximate the integral over $\lambda$ by a sum,
}
&= \lim_{P \to \infty}{
\frac{1}{P \partitionfuncbosons}
	\sum_{j=0}^{P-1}{
	\int{
    	\frac{1}{\fact{N}}\sum_{\sigma}{\langle \, \position_1,\ldots,\position_N \, | \, e^{-(\beta-\frac{\beta j}{P})\hamiltonian} \opA e^{-\frac{\beta j}{P} \hamiltonian} \opB \, | \, \position_{\sigma(1)}, \ldots, \position_{\sigma(N)} \rangle \, d\position_1 \ldots d\position_N}
    	}
    }
    }
\\
&= \lim_{P \to \infty}{
\frac{1}{P \partitionfuncbosons}
	\sum_{j=0}^{P-1}{
	\int{
    	\frac{1}{\fact{N}}\sum_{\sigma}{\langle \, \position_1,\ldots,\position_N \, | \, (e^{-\beta_P\hamiltonian})^{P-j} \opA (e^{-\beta_P \hamiltonian})^{j} \opB \, | \, \position_{\sigma(1)}, \ldots, \position_{\sigma(N)} \rangle \, d\position_1 \ldots d\position_N}
    	}
    }
    },
\\
\intertext{where $\beta_P = \beta/P$. We introduce a resolution of identity in position basis after each exponent to write
}
&= \lim_{P \to \infty}{
\frac{1}{P \partitionfuncbosons}
	\sum_{j=0}^{P-1}{
	\int{d\allbeadpositions \,
    	\frac{1}{\fact{N}}\sum_{\sigma}{
	    	\prod_{s=1}^{P-j}{\left(\langle \, \beadpos{1}{s},\ldots,\beadpos{N}{s} \, | \, e^{-\beta_P\hamiltonian} | \beadpos{1}{s+1},\ldots,\beadpos{N}{s+1} \rangle\right)}
	    	A(\beadpos{1}{P-j+1},\ldots,\beadpos{N}{P-j+1}) 
        }
    }
    }
    }
\\ \notag
&\times \prod_{s=P-j+1}^{P}{\left(\langle \, \beadpos{1}{s},\ldots,\beadpos{N}{s} \, | \, e^{-\beta_P\hamiltonian} | \beadpos{1}{s+1},\ldots,\beadpos{N}{s+1} \rangle\right)}
	    	B(\beadpos{1}{P+1},\ldots,\beadpos{N}{P+1}), 
\intertext{where $\allbeadpositions = \beadpos{1}{1},\ldots,\beadpos{1}{P},\ldots,\beadpos{N}{1},\ldots,\beadpos{N}{P}$,
and $\beadpos{\ell}{P+1}=\beadpos{\sigma(\ell)}{1}$, and we have used the fact that $\opA,\opB$ are diagonal in position-basis. Combining all the exponents to a single product, this is written more succinctly as}
&= \lim_{P \to \infty}{
\frac{1}{P \partitionfuncbosons}
	\sum_{j=0}^{P-1}{
	\int{d\allbeadpositions \,
    	\frac{1}{\fact{N}}\sum_{\sigma}{
    		A(\beadpos{1}{P-j+1},\ldots,\beadpos{N}{P-j+1}) 
	    	B(\beadpos{1}{P+1},\ldots,\beadpos{N}{P+1}) 
	    	\prod_{s=1}^{P}{\left(\langle \, \beadpos{1}{s},\ldots,\beadpos{N}{s} \, | \, e^{-\beta_P\hamiltonian} | \beadpos{1}{s+1},\ldots,\beadpos{N}{s+1} \rangle\right)}
    	}
    }
    }
    }
\\
\intertext{and changing the index which cycles through the imaginary-time slices $\beadpos{\cdot}{1},\ldots,\beadpos{\cdot}{P}$ we can write}
&= \lim_{P \to \infty}{
\frac{1}{P \partitionfuncbosons}
	\sum_{j=1}^{P}{
	\int{d\allbeadpositions \,
    	\frac{1}{\fact{N}}\sum_{\sigma}{
    		A(\beadpos{1}{j+1},\ldots,\beadpos{N}{j+1}) 
	    	B(\beadpos{1}{P+1},\ldots,\beadpos{N}{P+1}) 
	    	\prod_{s=1}^{P}{\left(\langle \, \beadpos{1}{s},\ldots,\beadpos{N}{s} \, | \, e^{-\beta_P\hamiltonian} | \beadpos{1}{s+1},\ldots,\beadpos{N}{s+1} \rangle\right)}
    	}
    }
    }
    }.
\\
\intertext{At this point we use the permutation-invariance of $\opA$ which implies that $A(\beadpos{1}{P+1},\ldots,\beadpos{N}{P+1}) = A(\beadpos{\sigma(1)}{1},\ldots,\beadpos{\sigma(N)}{1}) = A(\beadpos{1}{1},\ldots,\beadpos{N}{1})$, and likewise for $\opB$, to remove the dependence of the operators on the permutation and keep it only in the Boltzmann operators:}
&= \lim_{P \to \infty}{
\frac{1}{\partitionfuncbosons}
	\int{d\allbeadpositions \,
    		\left(\frac{1}{P}\sum_{j=1}^{P}{A(\beadpos{1}{j},\ldots,\beadpos{N}{j})}\right)
	    	B(\beadpos{1}{1},\ldots,\beadpos{N}{1}) 
	     \frac{1}{\fact{N}}\sum_{\sigma}{
	    	\prod_{s=1}^{P}{\left(\langle \, \beadpos{1}{s},\ldots,\beadpos{N}{s} \, | \, e^{-\beta_P\hamiltonian} | \beadpos{1}{s+1},\ldots,\beadpos{N}{s+1} \rangle\right)}
    	}
    }
    }.
\\
\intertext{Since $\beta_P \to 0$, we can apply the primitive Trotter approximation to each $e^{-\beta_P \hamiltonian}$, 
exactly as it is done when deriving the discretized path integral expression for static properties~\cite{tuckerman2010statistical}.
We find
}
&= \lim_{P \to \infty}{
\frac{1}{\partitionfuncbosons}
	\int{d\allbeadpositions \,
    		\left(\frac{1}{P}\sum_{j=1}^{P}{A(\beadpos{1}{j},\ldots,\beadpos{N}{j})}\right)
	    	B(\beadpos{1}{1},\ldots,\beadpos{N}{1}) 
	    C(P, \beta)\frac{1}{\fact{N}}\sum_{\sigma}{
	    	{e^{-\beta_P\left(\sum_{s=1}^{P}{\left(
	    		\sum_{\ell=1}^{N}{\springenergyprefix \left(\beadpos{\ell}{s+1}-\beadpos{\ell}{s}\right)^2} + V(\beadpos{1}{s},\ldots,\beadpos{N}{s})\right)}\right)}}
    	}
    }
    },
\\
\intertext{where $\springfrequency = P/{\beta \hbar}$ 
is the path integral spring frequency, and $C(P, \beta)$ is a constant pre-factor.
The same pre-factor also satisfies $\partitionfuncbosons = \lim_{P \to\infty} C(P,\beta) \partitionfuncpathintegralbosons$, where $\partitionfuncpathintegralbosons$ is the bosonic path integral partition function from Refs.~\citenum{hirshberg2019path,quadratic-pidmb} that was defined in the main text (up to a constant). We can thus rewrite
}
&= \lim_{P \to \infty}{
\frac{1}{\partitionfuncpathintegralbosons}
    \int{d\allbeadpositions d\allbeadmomenta \,
    		\left(\frac{1}{P}\sum_{j=1}^{P}{A(\beadpos{1}{j},\ldots,\beadpos{N}{j})}\right)
	    	B(\beadpos{1}{1},\ldots,\beadpos{N}{1})
	    	e^{-\beta_P \bosonshamiltonian}
    }
},
\\
\intertext{where $\allbeadmomenta$ indicates the momenta of all the beads $\beadmomentumbosons{1}{1},\ldots,\beadmomentumbosons{1}{P},\ldots,\beadmomentumbosons{N}{1},\ldots,\beadmomentumbosons{N}{P}$.
Finally, we replace the estimator $\left(\frac{1}{P}\sum_{j=1}^{P}{A(\beadpos{1}{j},\ldots,\beadpos{N}{j})}\right)
	    	B(\beadpos{1}{1},\ldots,\beadpos{N}{1})$ by 
its symmetrization over imaginary time slices:}
&= \lim_{P \to \infty}{
\frac{1}{\partitionfuncpathintegralbosons}
    \int{d\allbeadpositions d\allbeadmomenta \,
    		\left(\frac{1}{P}\sum_{j=1}^{P}{A(\beadpos{1}{j},\ldots,\beadpos{N}{j})}\right)
	    	\left(\frac{1}{P}\sum_{j=1}^{P}{B(\beadpos{1}{j},\ldots,\beadpos{N}{j})}\right)
	    	e^{-\beta_P \bosonshamiltonian}
    }
}.
\end{align}
This final symmetrization is an extension of the usual transformation from an estimator defined over only the first imaginary time-slice to an average of all them, as in distinguishable path integral molecular dynamics~\cite{tuckerman2010statistical}. 
            Such a transformation for bosons is implicit in Ref.~\citenum{hirshberg2019path} and will be made explicit in a forthcoming paper. The claim follows.

\subsection{Bosonic ring polymer correlations are even, symmetric functions}
\label{sec:even_symmetric_function}
We repeat the argument for distinguishable particles~\cite{craig_quantum_2004} to show that it remains unchanged for bosonic ring polymer correlation functions.
First we show that $\rpmdcorrbosons{A}{B}(-t) = \rpmdcorrbosons{A}{B}(t)$.
\begin{align}
\rpmdcorrbosons{A}{B}(-t) 
&= \ev{\rpmdopbosons{A}(0) \rpmdopbosons{B}(-t)}_{\bosonshamiltonian}
\\
\intertext{$\rpmdopbosons{B}(-t)$ corresponds to the backward-time evolution of the operator. 
By time-reversal symmetry, the same quantity should be obtained when inverting the momenta of all beads and instead evolving the system forward in time,
i.e., the time-reversal operation maps $(\allbeadpositions, \allbeadmomenta, -t) \rightarrow (\allbeadpositions, -\allbeadmomenta, t)$. However, since $\rpmdopbosons{B}$ (and likewise $\rpmdopbosons{A}$) depends only on position, and the RPMD Hamiltonian $\bosonshamiltonian(\allbeadpositions, \allbeadmomenta)$ is quadratic in momentum, both remain invariant under the time reversal operation. As a result, we obtain}
&=  \ev{\rpmdopbosons{A}(0) \rpmdopbosons{B}(t)}_{\bosonshamiltonian}
\\
&= \rpmdcorrbosons{A}{B}(t).
\end{align}

For symmetry w.r.t.\ the order of operators, we note on the other hand that
\begin{align}
\rpmdcorrbosons{A}{B}(-t) 
&= \ev{\rpmdopbosons{A}(0) \rpmdopbosons{B}(-t)}_{\bosonshamiltonian}
\\
\intertext{From the stationarity of the equilibrium distribution, we can shift time by $t$, and since classical observables commute,}
&= \ev{\rpmdopbosons{B}(0) \rpmdopbosons{A}(t)}_{\bosonshamiltonian}
\\
&=
\rpmdcorrbosons{B}{A}(t).
\end{align}
Notice here that it is important that the time evolution of $\opA,\opB$ is performed under the same Hamiltonian that is used in the Boltzmann operator.
Finally, since we have shown earlier that $\rpmdcorrbosons{A}{B}(t)  = \rpmdcorrbosons{A}{B}(-t)$, the claim follows.

\subsection{Bosonic harmonic limit}
\label{sec:bosonic_harmonic_limit}

The Hamiltonian is 
$\hamiltonian{} = \sum_{\ell=1}^{N}{\left(\frac{1}{2\mass} {\bf \hat{p}}^2_\ell + \frac{1}{2}\mass\omega^2 \opPosition_\ell^2\right)}$.
Consider the (classical) dynamics under the corresponding bosonic ring polymer Hamiltonian (\Cref{eq:bosons-ring-polymer-hamiltonian}) of the center-of-mass position
$\compos(t) = \frac{1}{N}\rpmdopbosons{\left(\sum_{\ell=1}^{N}{\position_\ell}\right)}(t) = \frac{1}{N}\frac{1}{P}
                \sum_{j=1}^{P}{
                    \sum_{\ell=1}^{N}{
                        \beadposbosons{\ell}{j}(t)
                    }
                }$.
The only change to the momentum of the center of mass
$\commomentum(t) = \frac{1}{N}\frac{1}{P}
                \sum_{j=1}^{P}{
                    \sum_{\ell=1}^{N}{
                        \beadmomentumbosons{\ell}{j}
                    }
                }$
is due to the external field:
\begin{equation}
\label{eq:harmonic-centroid-dynamics}
\frac{d}{dt} \commomentum(t)
= \frac{1}{N}\frac{1}{P}
    \sum_{j=1}^{P}{
        \sum_{\ell=1}^{N}{
            -\mass \omega^2 \beadposbosons{\ell}{j}
        }
    }
=
-\mass \omega^2 \mathbf{r}_c.
\end{equation}
Notice that this holds despite the sum over permutations which modifies the spring forces (and make the rings of different particles interact with each other).

According to~\Cref{eq:harmonic-centroid-dynamics}, the center of mass in the bosonic ring polymer Hamiltonian undergoes simple harmonic motion of frequency $\omega$ and mass $m$. The autocorrelation function of its position 
$\rpmdcorrbosons{\compos}{\compos}(t)$
is thus~\cite{craig_quantum_2004} a cosine function of $t$ with frequency $\omega$.
Since $\frac{1}{N} \rpmdcorrbosons{\sum_{\ell=1}^{N}{\position_\ell}}{ \sum_{\ell=1}^{N}{\position_\ell}}(t) = N \rpmdcorrbosons{\compos}{\compos}(t)$, 
this is true also for the ring polymer autocorrelation function that we are interested in.

It remains to show that the quantum Kubo-transformed autocorrelation 
$\frac{1}{N}\kubocorrbosons{\sum_{\ell=1}^{N}{\position_\ell}}{ \sum_{\ell=1}^{N}{\position_\ell}}(t)$ 
is also a cosine function of $t$ with frequency $\omega$, for then the short-time exactness (\Cref{sec:bosonic_short_time_limit}) guarantees that the correlations functions have the same amplitude and therefore entirely identical.
The operator $\sum_{\ell=1}^{N}{\hat{\position}_\ell}$ 
is symmetric under exchange, and thus the bosonic Kubo-transformed correlation function 
$\kubocorrbosons{\sum_{\ell=1}^{N}{\position_\ell}}{ \sum_{\ell=1}^{N}{\position_\ell}}(t)$ 
is an even function of time (\Cref{sec:bosonic_kubo_transformed_correlation_functions_symmetry_properties}), all we need to show that its only non-zero terms in the frequency domain are with frequencies $\pm \omega$. This is a consequence of the similar property of distinguishable Kubo-transformed correlation functions.

Expanding in the basis formed from the symmetrization of the basis that is the product of the harmonic oscillator energy states, one per particle,
\begin{align}
	\kubocorrbosons{A}{B}(t) 
	&\propto \int_{0}^{\beta}{
    d\lambda \, 
    \tracebosons\left\{
    e^{-(\beta-\lambda)\hamiltonian} \opA e^{-\lambda \hamiltonian} e^{+it\hamiltonian/\hbar} \opB e^{-it\hamiltonian/\hbar}
    \right\}
    }
	\\
	&= \int_{0}^{\beta}{d\lambda \, \frac{1}{\fact{N}}\sum_{\sigma}{\sum_{n_1,\ldots,n_N}{\left\langle n_1,\ldots,n_N \, \big{|} \, e^{-(\beta-\lambda)\hamiltonian} \opA e^{-\lambda \hamiltonian} e^{+it\hamiltonian/\hbar} \opB e^{-it\hamiltonian/\hbar} \, \big{|} \, n_{\sigma(1)}, \ldots, n_{\sigma(N)} \right\rangle}}}
	\\
	\intertext{Since $\hamiltonian$ is separable in the different particles, and each $n_\ell$ is an eigenstate of the corresponding Hamiltonian (denote by $E_{n_\ell}$ the corresponding eigenvalue),}
	&= \int_{0}^{\beta}{d\lambda \, \frac{1}{\fact{N}}\sum_{\sigma}{\sum_{n_1,\ldots,n_N}{e^{-(\beta-\lambda)(\sum_{\ell}{E_{n_\ell}})} \left\langle n_1,\ldots,n_N \, \big{|} \, \opA e^{-\lambda \hamiltonian} e^{+it\hamiltonian/\hbar} \opB \, \big{|} \, n_{\sigma(1)}, \ldots, n_{\sigma(N)} \right\rangle e^{-it(\sum_{\ell}{E_{n_\ell}})/\hbar}}}}
	\\
	\intertext{Introducing a resolution of identity through a sum over the same energy basis,}
	&= \begin{aligned}
	    \int_{0}^{\beta}d\lambda \, \frac{1}{\fact{N}}\sum_{\sigma}\sum_{n_1,\ldots,n_N}\sum_{m_1,\ldots,m_N}e^{-(\beta-\lambda)(\sum_{\ell}{E_{n_\ell}})} \left\langle n_1,\ldots,n_N \, \big{|} \, \opA \, \big{|} \, m_1,\ldots,m_N 
        \right\rangle e^{-\lambda (\sum_{\ell}{E_{m_\ell}})} e^{+it(\sum_{\ell}{E_{m_\ell}})/\hbar}
        \\
        \cdot \left \langle m_1,\ldots,m_N \, \big{|} \, \opB \, \big{|} \, n_{\sigma(1)}, \ldots, n_{\sigma(N)} \right\rangle e^{-it(\sum_{\ell}{E_{\ell}{n_\ell}})/\hbar}
        \end{aligned}
	\\
	&= \begin{aligned}
	\int_{0}^{\beta}d\lambda \, \frac{1}{\fact{N}}\sum_{\sigma}\sum_{n_1,\ldots,n_N}\sum_{m_1,\ldots,m_N}\left\langle n_1,\ldots,n_N \, \big{|} \, \opA \, \big{|} \, m_1,\ldots,m_N \right\rangle \left \langle m_1,\ldots,m_N \, \big{|} \, \opB \, \big{|} \, n_{\sigma(1)}, \ldots, n_{\sigma(N)} \right\rangle 
    \\
    \cdot e^{-(\beta-\lambda)(\sum_{\ell}{E_{n_\ell}})} e^{-\lambda (\sum_{\ell}{E_{m_\ell}})} e^{-it(\sum_{\ell}{(E_{n_\ell}-E_{m_\ell})})/\hbar}
    \end{aligned}
\end{align}
Thus the Fourier transform of $\kubocorrbosons{A}{B}(t)$ is non-zero at frequencies $\sum_{\ell}{(E_{n_\ell}-E_{m_\ell})}/\hbar$ for energy states $n_1,\ldots,n_N$ and $m_1,\ldots,m_N$ such that $\left\langle n_1,\ldots,n_N \, \big{|} \, \opA \, \big{|} \, m_1,\ldots,m_N \right\rangle \neq 0$ and $\left \langle m_1,\ldots,m_N \, \big{|} \, \opB \, \big{|} \, n_{\sigma(1)}, \ldots, n_{\sigma(N)} \right\rangle \neq 0$ for some permutation $\sigma$.

For simplicity, consider a one-dimensional system. In our case, $\opA = \opB = \sum_{\ell}{\hat{\positiononedim}_\ell}$. Thus
\begin{align*}
	\left\langle n_1,\ldots,n_N \, \big{|} \, \opA \, \big{|} \, m_1,\ldots,m_N \right\rangle
	&=
	\sum_{\ell}{\left\langle n_1,\ldots,n_N \, \big{|} \, \hat{\positiononedim}_\ell \, \big{|} \, m_1,\ldots,m_N \right\rangle}
	\\
	&=
	\sum_{\ell}{\left\langle n_\ell \, \big{|} \, \hat{\positiononedim}_\ell \, \big{|} \, m_\ell \right\rangle \prod_{i\neq \ell}{\delta(n_i - m_i)}}
\end{align*}
In an harmonic oscillator, $\left\langle n_\ell \, \big{|} \, \hat{\positiononedim}_\ell \, \big{|} \, m_\ell \right\rangle \neq 0$ only if $n_\ell = m_\ell \pm 1$. Thus $\left\langle n_1,\ldots,n_N \, \big{|} \, \opA \, \big{|} \, m_1,\ldots,m_N \right\rangle$ is non-zero only if there is one $\ell$ such that $n_i = m_i$ for every $i \neq \ell$ and $n_\ell = m_\ell \pm 1$, which in an harmonic oscillator means that $E_{n_\ell} = E_{m_\ell} \pm \hbar \omega$. This means that $\left\langle n_1,\ldots,n_N \, \big{|} \, \opA \, \big{|} \, m_1,\ldots,m_N \right\rangle \neq 0$ only if $\sum_{\ell}{(E_{n_\ell}-E_{m_\ell})}/\hbar = \pm \omega$. The claim follows.

\section{Simulation details}
\label{sec:simulation-details}

Details of the NVT simulation:
\begin{table}[htbp]
\centering
\label{tab:sho_sim_details}
\begin{tabular}{cccc}
\hline
\hline
\vspace{1.25mm}
Harmonic \\
\hline
$\beta$ [a.u.] 
& 
Timestep [fs] 
& 
Beads 
& 
Steps [$10^7$] 
\\
\hline
 8.0 & 0.5 & 32 & 4 \\
\hline
\hline
\vspace{1.25mm}
Weakly anharmonic \\
\hline
$\beta$ [a.u.] & Timestep [as] & Beads & Steps [$10^7$] \\
\hline
 1.0 & 0.05 & 16 & 4 \\
\hline
 8.0 & 0.1 & 32 & 4 \\
\hline
\hline
\vspace{1.25mm}
Strongly anharmonic \\
\hline
$\beta$ [a.u.] & Timestep [as] & Beads & Steps [$10^7$] \\
\hline
 1.0 & 0.1 & 16 & 2 \\
\hline
 8.0 & 0.1 & 32 & 2 \\
\hline
\hline
\vspace{1.25mm}
Lieb-Liniger (Gaussian) \\
\hline
$T$ [K] & Timestep [fs] & Beads & Steps [$10^7$] \\
\hline
 1.0 & 2.0 & 84 & 2 \\
\hline
\end{tabular}
\end{table}

The RPMD correlation function is obtained by averaging over $4000-8000$ trajectories. The initial conditions for these trajectories are sampled from the NVT trajectory separated by equal time intervals. All the simulations are performed in atomic units (a.u.).

\section{Bosons vs. Distinguishable Particles: Position autocorrelation function}
\label{comparison-boson-dist-modelpotential}
\Cref{fig:Cxx_comparison} shows the comparison between the Kubo-transformed real-time position autocorrelation functions for bosons and distinguishable particles. Since bosonic RPMD incorporates the bosonic exchange symmetry, the bosonic RPMD correlation function captures the dynamical signature of the exchange symmetry on the correlation functions.

\begin{figure}[!ht]
\centering
\includegraphics[width=1\textwidth]{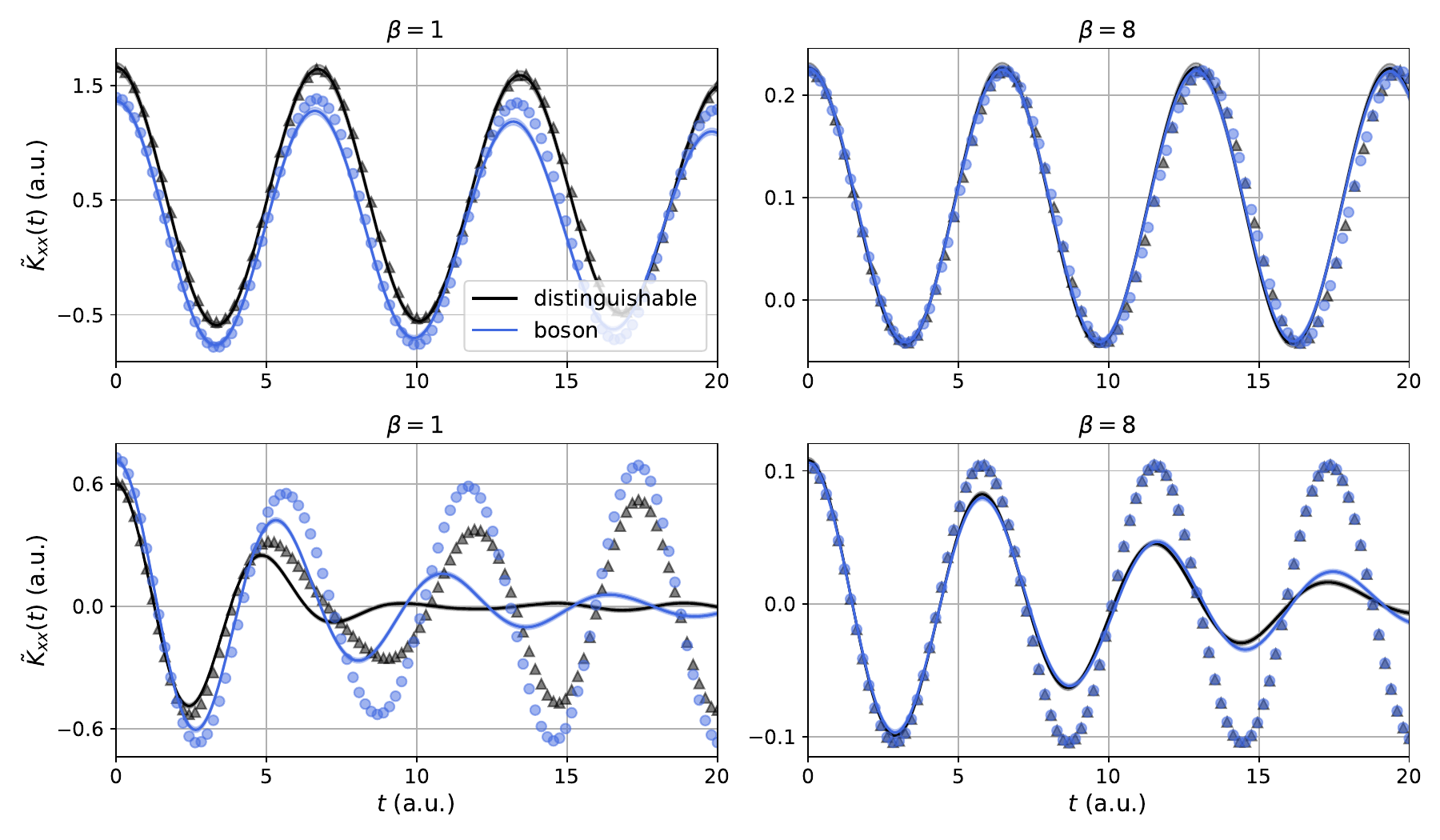}
\caption{Real-time Kubo-transformed position autocorrelation function for non-interacting ($N=4$) bosons (blue) and distinguishable particles (black) at two different temperatures in a weakly (upper panel) and strongly anharmonic potential (lower panel). Lines denote bosonic and distinguishable RPMD results; symbols denote the numerically exact results (bosons: blue circles; distinguishable particles: black triangles).}
\label{fig:Cxx_comparison}
\end{figure}

\section{Parameters for a 1D system of interacting bosons}
\label{sec:parameters-mLL}

A general two-body Hamiltonian of the form
\begin{equation}
    \hat{H} 
    = 
    -\frac{\hbar^2}{2m_1} \pdv[2]{x_1}
    -\frac{\hbar^2}{2m_2} \pdv[2]{x_2}
    + 
    V
    \left(
        \left|
            x_1 - x_2
        \right|
    \right),
\end{equation}
can be written in terms of the center of mass $R=\frac{m_{1}x_{1}+m_{2}x_{2}}{m_{1}+m_{2}}$ and the relative coordinate $r = x_1 - x_2$ as
\begin{equation}
    \hat{H} 
    = 
    -\frac{\hbar^{2}}{2M} \pdv[2]{R}
    -
    \frac{\hbar^{2}}{2\mu} \pdv[2]{r}
    +
    V\left(r\right),
\end{equation}
where $M = m_1 + m_2$ and $\mu = \frac{m_1 m_2}{m_1 + m_2}$ are the total and reduced masses, respectively. For bosons with Gaussian pair interaction, $m_1 = m_2 = m$ and $V(r)= \frac{g}{\sqrt{2\pi \sigma_g^2}}e^{-\frac{r^2}{2\sigma_g^2}}$. Solving the  Schr\"{o}dinger equation corresponding to the relative motion part of the Hamiltonian yields the following approximate expression for the $s$-wave scattering length:~\cite{jeszenszki2018s}
\begin{equation}
a \approx \frac{2\sigma_g}{\sqrt{\pi}} - \frac{2 \hbar^2}{mg}.
\end{equation}

To approximate the delta interaction with the Gaussian interaction, we equate the scattering lengths for both potentials.
\begin{equation}\label{eq:scattering_length}
\frac{2\sigma_g}{\sqrt{\pi}} - \frac{2 \hbar^2}{mg} = -\frac{2\hbar^2}{mg_{\text{LL}}}.
\end{equation}

For a given $g_{\text{LL}}$,~\Cref{eq:scattering_length} provides infinitely many possible pairs $\set{g, \sigma_g}$. We fix the optimal $\set{g, \sigma_g}$ by comparing the average energy of the Gaussian interacting potential with that of the Lieb-Liniger system for $N=2$.~\Cref{fig:LL-N2} shows the comparison of the average energy across temperatures for the optimal $\set{g, \sigma_g}$.

\begin{figure}[ht]
\centering
\includegraphics[width=0.6\textwidth]{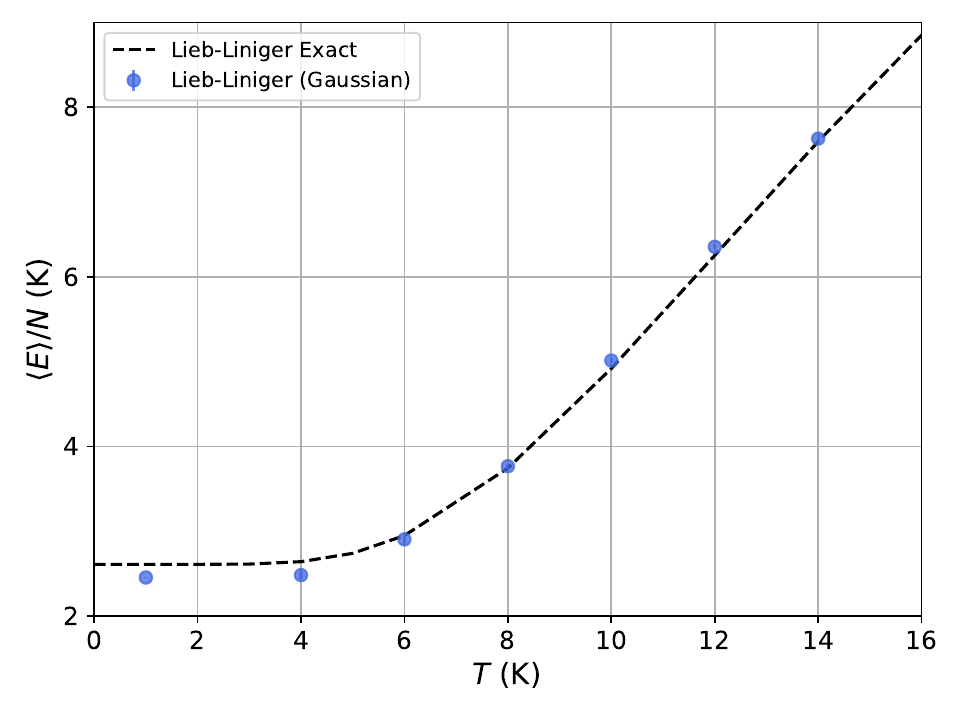}
\caption{Comparison between the average energy of the Lieb-Liniger model ($g_{\text{LL}} = \unit{10}{K\AA}$) and that of the bosons with the pairwise gaussian interaction (Optimal parameters: $g = \unit{6}{K\AA}$, $\sigma_g = \length{0.118234}$) for $N=2$.}
\label{fig:LL-N2}
\end{figure}

\section{Bosons vs. Distinguishable Particles: RPMD density-density correlation function and Dynamic Structure Factor}
\label{sec:comparison-boson-dist-LL}
\Cref{fig:isf_dsf_comparison} compares the real-time RPMD density-density correlation function for bosons and distinguishable particles at different momentum transfers for the system described by the Hamiltonian in \Cref{eq:modified-LL-ham}. Bosonic RPMD captures the dynamical signature of the exchange symmetry on the RPMD correlation function. \Cref{fig:isf_dsf_comparison} also shows the DSF for $k/k_F = 0.125$ for both bosons and distinguishable particles. Note that the peak at $\omega = 0$ is absent for the distinguishable particles.
\begin{figure}[ht]
\centering
\includegraphics[width=1\textwidth]{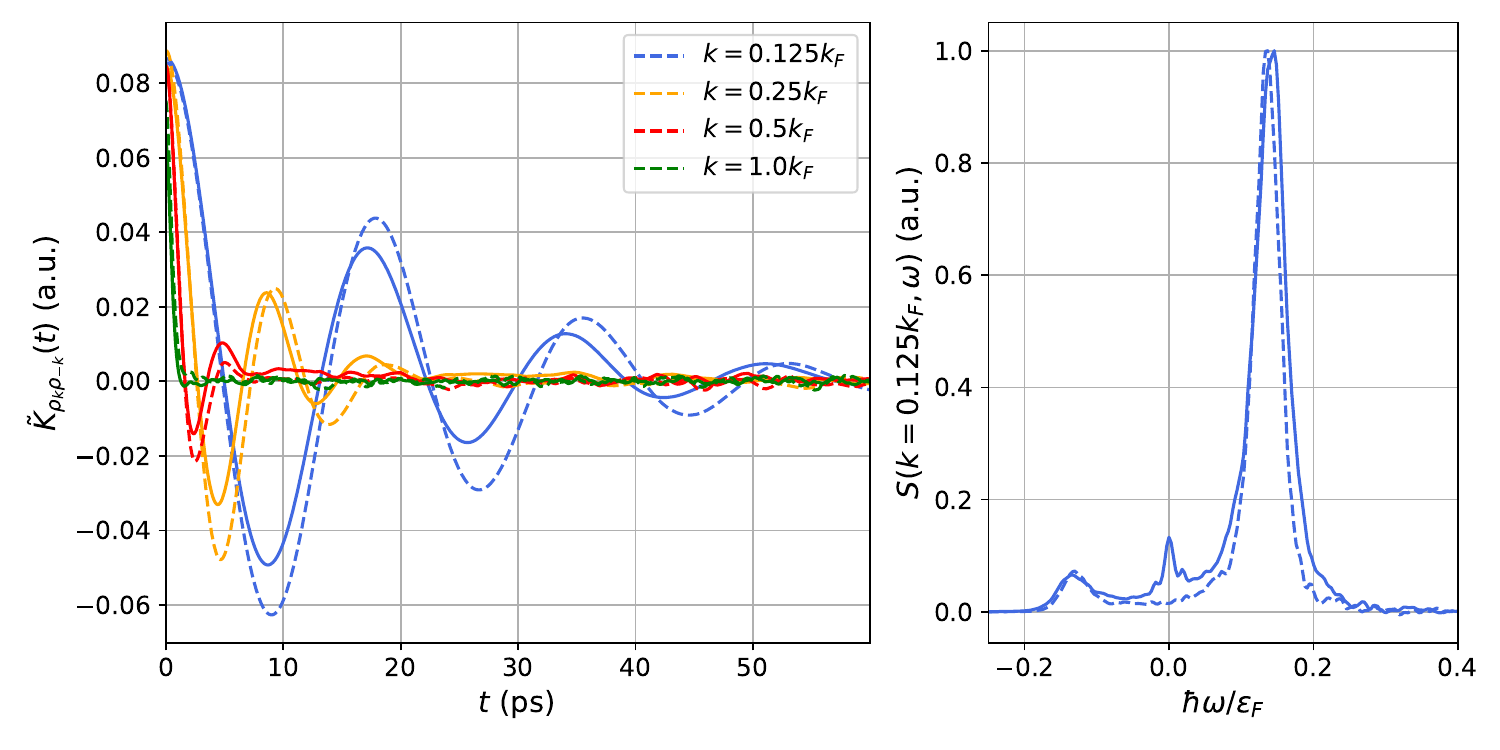}
\caption{(Left panel) Comparison between the real-time RPMD density-density correlation functions for bosons and distinguishable particles at four different momentum transfers $k / k_F = 0.125, 0.25, 0.5, 1.0$ at $T = \unit{1}{K}$. (Right panel) Comparison between the DSFs (scaled) for bosons and distinguishable particles at momentum transfer $k / k_F = 0.125$ at $T = \unit{1}{K}$ ($\varepsilon_F = \hbar^2 k_F^2/2m$). Solid and dashed lines represent the bosons and the distinguishable particles, respectively.}
\label{fig:isf_dsf_comparison}
\end{figure}

\section{Noise suppression in the dynamic structure factor}
\Cref{fig:isf_dsf_noise} shows the real-time RPMD density-density correlation functions (with and without the exponential filter) and the DSFs obtained from them. The real-time correlation function shows fluctuations at long times. Because of these long time fluctuations, the DSF obtained from them becomes very noisy, particularly for large momentum transfers ($k/k_F = 0.5, 1.0$). 
To mitigate this, the correlation function is multiplied by an exponential window function $\exp(-t/\tau)$, which suppresses the fluctuations at long times, while leaving the dominant short time behavior unaffected. As verified in~\Cref{fig:isf_dsf_noise}, this procedure does not alter the peak location or the overall features of the DSF. For the four different momentum transfers $k/k_F = 0.125, 0.25, 0.5, 1.0$, the damping constants used are $\tau \approx \unit{480}{ps}, \unit{240}{ps}, \unit{48}{ps}, \unit{24}{ps}$, respectively.

\label{sec:noise_suppresion_dsf}
\begin{figure}[ht]
\centering
\includegraphics[width=1\textwidth]{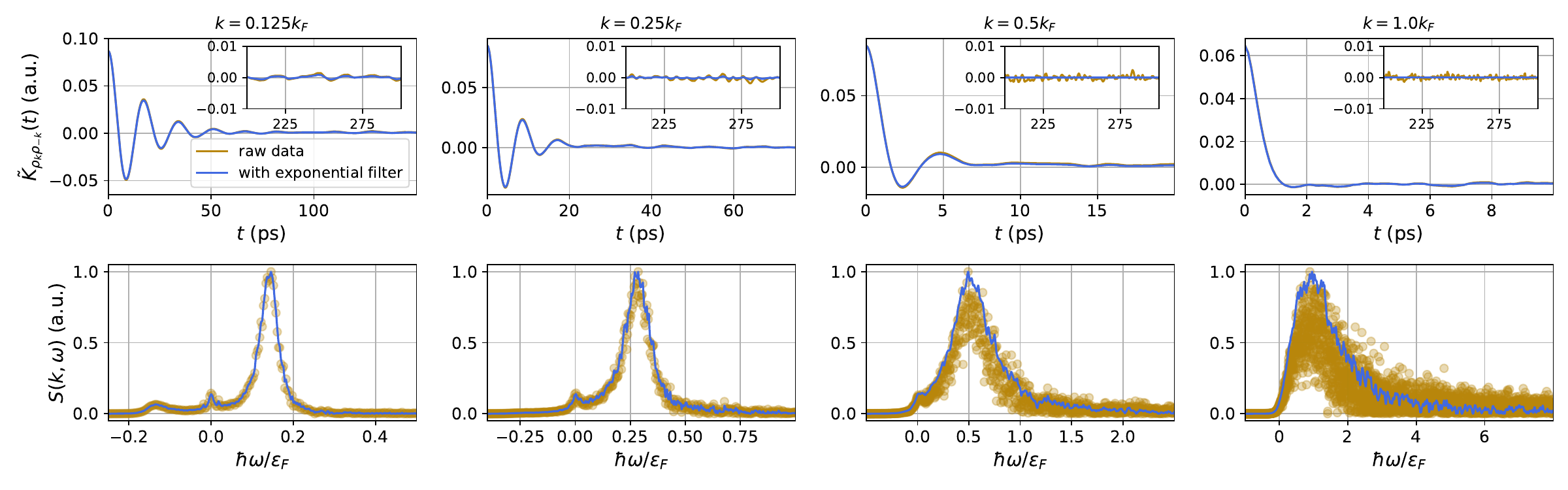}
\caption{(Top panel) Bosonic RPMD density-density real-time correlation functions without and with the exponential filter applied at four different momentum transfers $k / k_F = 0.125, 0.25, 0.5, 1.0$ at $T = \unit{1}{K}$. Inset shows the fluctuations at long times. (Bottom panel) Comparison between the DSFs computed from the RPMD density-density correlation function without and with the exponential filter applied ($\varepsilon_F = \hbar^2 k_F^2/2m$).}
\label{fig:isf_dsf_noise}
\end{figure}

\section{Bosonic RPMD density-density correlation function at different temperatures}
\label{sec:isf_diff_Ts}
\Cref{fig:isf_diff_Ts} shows the bosonic real-time density-density correlation function for the system described by the Hamiltonian in \Cref{eq:modified-LL-ham} at different temperatures $T = \unit{4}{K}$, $\unit{8}{K}$, and $\unit{14}{K}$.

\begin{figure}[ht]
\centering
\includegraphics[width=1\textwidth]{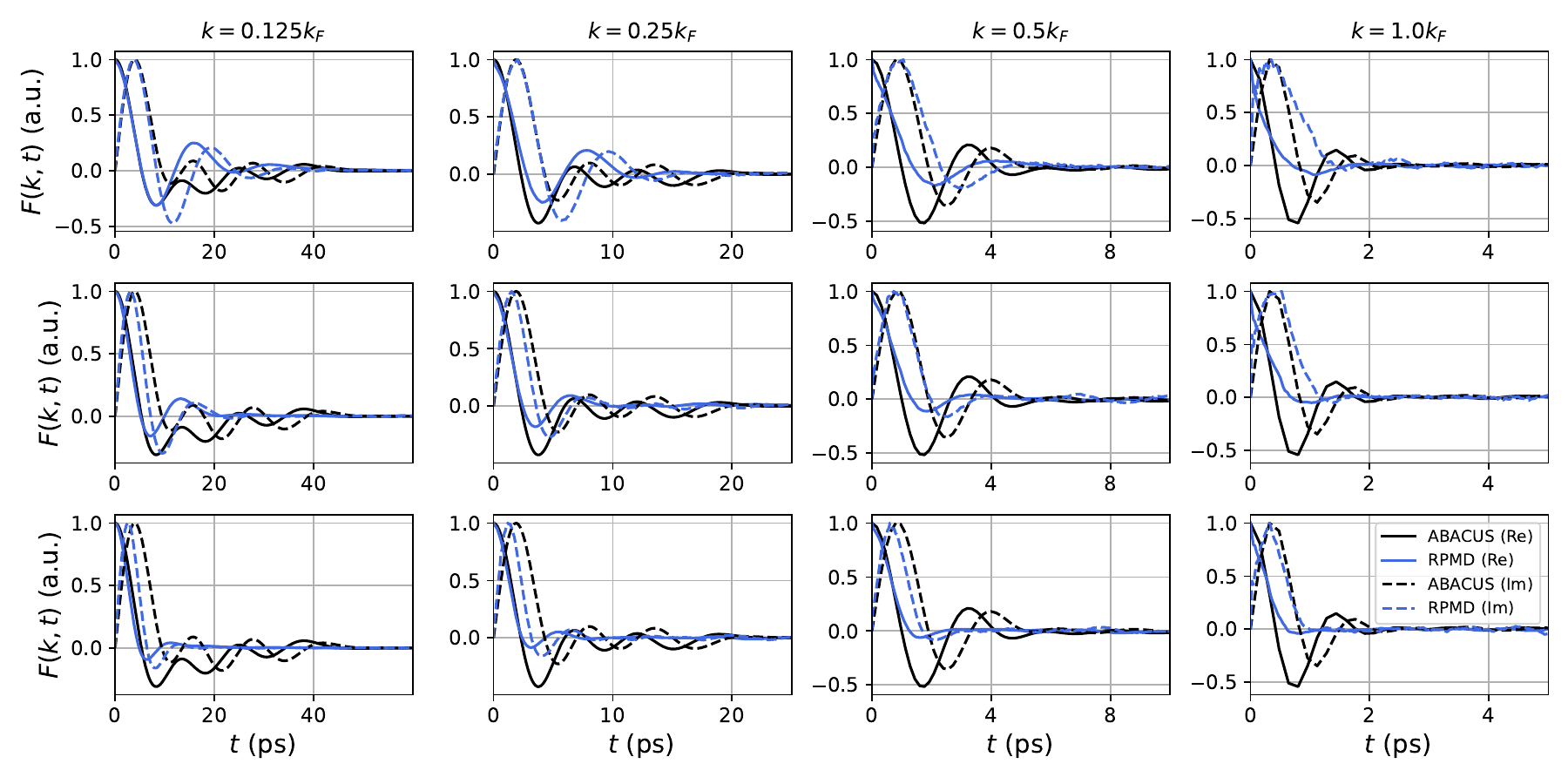}
\caption{Intermediate scattering function, or the real-time density–density correlation function (scaled) at different temperatures: $T = \unit{4}{K}$ (upper panel), $\unit{8}{K}$ (middle panel), and $\unit{14}{K}$ (lower panel), for different momentum transfers $k / k_F = 0.125, 0.25, 0.5, 1.0$. System description: $N = 32$, $L = \length{16}$.}
\label{fig:isf_diff_Ts}
\end{figure}
\fi

\end{document}